\documentclass[aps,pra,showpacs,twoside,twocolumn,10pt]{revtex4-1}
\usepackage[colorlinks=true, citecolor=red, urlcolor=blue ]{hyperref}
\usepackage{epsfig,newlfont,amssymb,amsfonts,amsmath,bm,subfigure,palatino,mathtools,amsthm,braket,soul,enumitem,color,graphics,graphicx,times,physics,bbold}
\usepackage[normalem]{ulem}
\usepackage{xcolor}
\usepackage{physics}
\usepackage{dsfont}
\usepackage{float}
\usepackage{amsmath}
\usepackage{tabularx}
\usepackage{array}
\usepackage[export]{adjustbox}
\usepackage{mathtools}

\newcolumntype{Y}{>{\centering\arraybackslash}X}

\begin{document}
\title{
Classical correlations for Generic States are Fragile under Decoherence
}

\author{Tanoy Kanti Konar, Arghya Maity, Aditi Sen(De)}
\affiliation{Harish-Chandra Research Institute, A CI of Homi Bhabha National Institute,  Chhatnag Road, Jhunsi, Allahabad - 211019, India}

\begin{abstract}
Quantum correlations typically decrease with increasing noise, although  classical correlators (CCors)  may rise for a particular class of states with noise. To analyse the behavior of classical correlation (CC) in the presence of local noise, we scrutinize the set of classical correlators, axiomatic CC measures like classical discord, and local work for Haar uniformly generated states.  Like quantum correlation measures, we illustrate that when noise levels rise, the average value of the CC measures for noisy output states obtained from random input states  decreases for most of the channels. We also demonstrate a connection between the CCors of the noise-affected multipartite states that are produced and the CCors of the initial states that exhibit exponential, polynomial, and constant behavior as the noise level changes. Moreover, based on CCors of the generalised $N$-qubit $W$ state as input, we determine a method to discriminate between the quantum channels, namely phase damping, depolarizing, and amplitude damping channels. We also relate classical, quantum, and total correlation measures that exhibit a comparable reaction to decoherence for generic states.
\end{abstract}

\maketitle

\section{Introduction}
\label{sec:intro}

In a quantum world, an isolated system is hard to prepare since they are always interacting with their surroundings. Therefore, it is crucial to investigate how the environment impacts the physical properties of a quantum system and in turn, how decoherence affects the efficiency of quantum information processing tasks. It is known that quantum correlations (QC), whether they take the form of entanglement \cite{Horodecki_RMP_2009} or measure independent of entanglement \cite{Kmodi_RMP_2012,bera2017,Alexander-Streltsov_book,Fanchini_book} typically decrease with an increase in noise, demonstrating their fragility to decoherence. The sudden death of entanglement with the increase of noise is one of the intriguing phenomena that has been reported \cite{Zyczkowski_PRA_2001,Halliwell_PRA_2004,bruno2008,ting2009,ting_OC_2006}. However, other QC measures like quantum discord \cite{ollivier2001,Kmodi_RMP_2012, bera2017}, and quantum work deficit \cite{horodecki2002,micha2003,aditi2005} for a certain class of states remain constant before degrading under various types of noise in the system, a phenomenon known as freezing of QC \cite{mazzola2010,benjamin2013,chanda2015,bera2017}.

On the other hand, classical correlators (CCors) are essential in many domains of physics including  comprehending the quantum phase transition in many-body physics \cite{sachdev_2011},  determining whether local realism is violated using correlation function Bell inequalities \cite{JSbell_PPF_1964,Clauser_PRL_1969,horodechki1995} which are also employed to identify the entangled state in experiments, and to assess the utility of states in teleportation  \cite{horodecki1996a}. Hence, understanding their behavior is again crucial when the system interacts with the environment although unlike QCs, investigations in this direction is quite limited, even in  bipartite scenarios. This is due to the fact that although the characteristics of a valid quantum correlation measures which include  entanglement, quantum discord and work-deficit \cite{Kmodi_RMP_2012,bera2017}, measurement induced nonlocality \cite{Luo_PRL_2011} are well established,  the definition of  classical correlations in a multipartite system is not fully understood.

\begin{figure}
		\includegraphics[width=\linewidth]{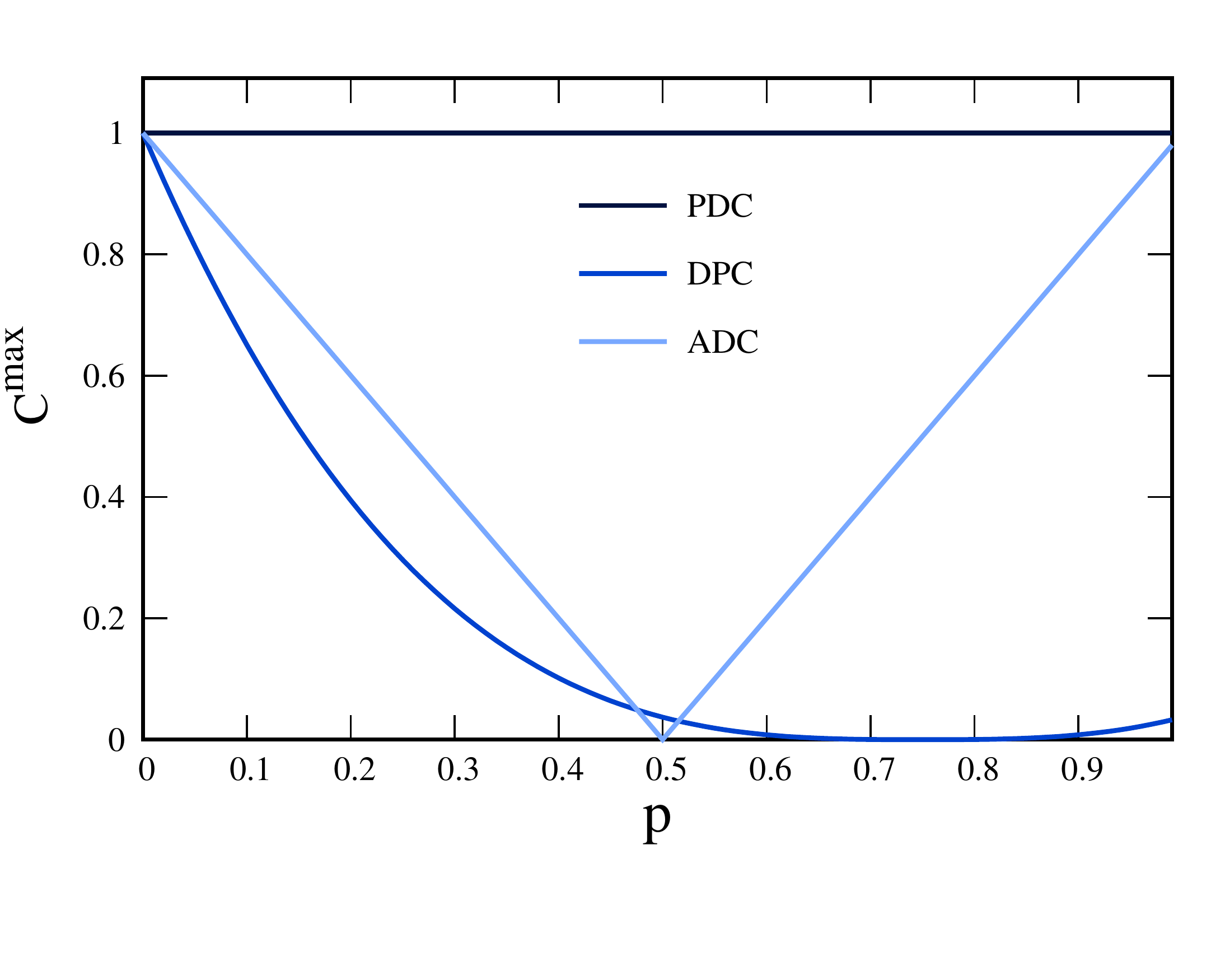}
		\caption{\textbf{\(C^{\max}= \max [C_{xxx}, C_{yyy}, C_{zzz}]\) (ordinate))  against noise strength, \(p\) (abscissa).}  The  initial state is the three-qubit \(W\) state, given by \(|W^3\rangle = \frac{1}{\sqrt{3}} (|001\rangle + |010\rangle + |100\rangle)\). It can be easily found that \(C^{\max}\) is always attained by \(C_{zzz}\) and hence ordinate can also be replaced by \(C_{zzz}\).   PDC, DPC and ADC represent phase damping, depolarizing and amplitude damping channels respectively. Unlike QCs, \(C^{\max}\) can increase with the increase of noise which acts on individual qubit of the initial state. It is interesting to find whether such behavior is true for generic states or not.  Both the axes are dimensionless. }
		    \label{fig:gen_w_cmax}
\end{figure}

In this work, our primary objective  is to examine how the introduction of noise alters the behavior of CCs. For analysis, we consider genuine classical correlators, a set of classical correlators made up of all possible classical correlators obtained from different reduced states  as well as the total multipartite states and the distribution of classical correlation measures based on known measures in a bipartite regime.
The entire study has yet another significant component. Classical correlation measures can behave in distinct ways from QCs  for certain classes of states. For instance, unlike entanglement, there is a region of noise where CCs grows with the increase in noise when all the three-qubits of a $W$ state \cite{Dur_PRA_2000} are influenced by noise (amplitude damping and depolarising noise) (see Fig. \ref{fig:gen_w_cmax}). As a result, one would anticipate that CC will respond differently in a noisy environment than QC measures. However, we demonstrate that this is not the case by establishing a connection between classical correlators obtained after the action of local noise and that of prior noise which exhibits  exponential or polynomial decay of CCors in most situations. Moreover, it turns out that the decay rate of classical correlations of a particular class of states can differentiate between local channels on states.

To check the response of noise on CC measures for generic states, we also Haar uniformly generate multipartite states with three-, four- and five-qubits and analyze the evolution of distributed CC measures and genuine classical correlators with noise and the number of parties. This is in the hopes that recent work has shown that, even though  CC measures do not obey the monogamy relation \cite{Ekert_PRL_1991,Bennett_PRA_1996,Coffman_PRA_2000,BM_Terhal_IBM_2004,Osborne_PRL_2006,Gour_CP_2012,Streltsov_PRL_2012,prabhu2012,monorev}, they do have a constraint on the sharing of CCs when random  multipartite states are considered. Moreover, studying randomly generated states is crucial for identifying the typical characteristics of quantum states \cite{Benjamin_PRA_1996,bengtsson_zyczkowski_2006,Patrick_CMP_2006,Gross_PRL_2009,Hastings_NaturePhysics_2009,Winter_PRL_2009,Soorya_Aditi_PRA_2019,Mahasweta19,Waldemar_PRA_2019,Gupta'21}.

We report that  CC measures obtained from generic states as inputs decay as noise level rise, irrespective of the channels and with the number of parties. The frequency distribution of CC measures for random states demonstrate that, similar to QC measures, the average shareability of CC measures decreases with noise before vanishing in high noise while the standard deviation decreases except for some anomaly found for amplitude damping channel. Furthermore, we discover that although average entanglement of the Haar uniformly generated states increases with the increase in the number of parties \cite{Winter_PRL_2009, eisert2009, Soorya_Aditi_PRA_2019}, the average distributed CCs decreases even in presence of noise. The depolarising channel (DPC) exhibits a more marked similarity in decay behavior between quantum and classical correlation than amplitude (AD) and phase damping channels (PDC). In addition, we also demonstrate that for a fixed distributed total correlations \cite{henderson2001, Winter_PRA_2005}, there are lower and upper bounds on distributed CC measures in the definition of quantum discord and the optimal distributed CC values are connected linearly with distributed total correlations which are not the case for other CC measures. 

This paper is organised in the following way. In Sec. \ref{sec:Figmerits}, we present the figures of merits that we consider to quantify CCs for multipartite states. We obtain the relation of the set of classical correlators between the initial and the evolved noisy states for three different channels in Sec. \ref{sec:CCnoise} while in Sec. \ref{subsec:discrimination}, we provide a method based on CCors of the $W$ state to distinguish between three different channels.   The normalised frequency distribution of different classical correlation measures for Haar uniformly generated multiqubit pure states as inputs and the behavior of statistical quantities obtained from the  distributions in presence of noise  are analyzed in Sec. \ref{sec:frequency}. Sec. \ref{sec:CCvsQCTC} connects the patterns  of  CCs with QCs and total correlations when states are sent through noisy channels. We summarize our results in  Sec. \ref{sec:conclusion}.

\section{Figures of merits for Correlations}
\label{sec:Figmerits}

Quantification of quantum correlations, especially for pure states, is relatively better understood compared to other correlation present in states. However, measures for classical correlation and total correlation which comprises both classical and quantum correlation were introduced  for bipartite states. Based on such characterization and motivatio from many-body physics, we introduce correlation measures for multipartite states.

{\it Genuine and  set of classical correlators. } The traditional genuine classical correlators  are useful in several areas of physics like in derivation of Bell inequalities \cite{brunner2014,horodechki1995},  in detecting quantum phase transition in spin models \cite{sachdev_2011}.  For an arbitrary \(N\)-party state, \(\rho_{B_1\ldots B_N}\), they can be defined as (see Appendix \ref{cc_measures}) 
\begin{eqnarray}
C'_{j_{1}\ldots j_{N}} = \tr ( \sigma_{j_1} \otimes \ldots \otimes \sigma_{j_{N}} \rho_{B_1\ldots B_N}),  j_k=x,y,z,
\label{correlator}
\end{eqnarray}
where \(\sigma_{j_k}\) (\(k=1, \ldots, N\)) are any Pauli matrices at any individual site, $B_k$. Since \(C'_{j_{1}\ldots j_{N}}\) can take any values between \(-1\) and \(1\), we  rescale classical correlators as \(C_{j_{1}\ldots j_{N}} = |C'_{j_{1}\ldots j_{N}}|\), so that \( 0\le C_{j_{1}\ldots j_{N}}\le 1\). When one or more \(\sigma_{j_k}\) is identity, we call the CCors to be non-genuine. Alternatively, non-genuine CCors can  be defined using reduced density matrices, i.e., \(C_{j_1\ldots j_k}=\tr(\sigma_{j_1} \otimes \ldots \otimes \sigma_{j_{k}} \rho_{B_1\ldots B_k})\) (\(k < N\)) where \(\rho_{B_1\ldots B_k}=\tr_{\Bar{B_k}}\rho_{B_1\ldots B_N}\) is $k$-party reduced density matrix of $\rho_{B_1\ldots B_N}$, obtained after tracing out \(N-k\) parties. We can define genuine maximal CCors as
\begin{equation}
    C^{\max}_N=\underset{j_k}{\max}\left[C_{j_{1}\ldots j_{k}}\right].
    \label{eq::gen_max}
\end{equation}
Similarly, for fixed reduced states of \(k\)-parties, we can define maximum non-genuine \(k\)-partite CCors as
\begin{equation}
    C_k^{\max}=\underset{j_k}{\max}\left [ C_{j_{1}\ldots j_{k}}\right], k<N.
    \label{eq::nongen_max}
\end{equation}
Since for a given state \(\rho_{B_1\ldots B_N}\), there can be many possibilities to obtain \(k\)-party reduced states, we can have a set, \(\{C_k^{\max}\}\)  (\(k=1,\ldots N\)), i.e.,
\begin{eqnarray}
\mathcal{S}^{CC} = \{ \{C_{1}^{\max}\}, \{C_{2}^{\max}\}, \cdots, C^{\max}_N\},
\end{eqnarray}
where \(\{C^{\max}_{k}\}\) represents the set of maximal classical correlators obtained from all possible \(k\)-party reduced states of a \(N\)-party state, defined in Eq. (\ref{eq::nongen_max}). Note that in the case of symmetric states, all reduced states coincide and hence each set consists of a single element. Since $\mathcal{S}^{CC}$ contains all types of CCors present in a $N$-party state, it can  capture the classical correlation possessed by states. The aim of the paper is to study the effects of noisy channels on \(\mathcal{S}^{CC}\) or its elements individually. For  simplicity, we denote the maximum genuine CCor as \(C^{\max}\) by omitting the subscript. In this paper, we mostly compute \(C^{\max}\) by considering all \(\sigma_{j_k}\)s to be the same.

{\it Distributed correlations. } Correlations -- both classical and quantum -- are well established for bipartite states. For instance, 
there are axiomatic classical correlation measures introduced for bipartite states \cite{henderson2001, aditi2005}. These measures can be adopted to quantify how correlations are distributed (shared) among parties in a multipartite system. For a given \(N\)-party state, \(\rho_{B_1\ldots B_N}\), and for a fixed correlation measure, \(\chi\), distributed correlations among \(N\) parties can be quantified as 
\begin{eqnarray}
\mathcal{D}^{\chi} (\rho_{B_1\ldots B_N}) = \sum_{i=2}^N \chi(\rho_{B_1 B_i}),
\label{eq::distributed}
\end{eqnarray}
where \(\rho_{B_1 B_i}\) represents the reduced density matrices of \(\rho_{B_1\ldots B_N}\) with $B_1$ being the nodal (central) observer.  Correlation measures can be classical correlation measures  like maximum CCors, CC associated with QD which we call classical discord denoted by \(\mathcal{CD}\) \cite{henderson2001,ollivier2001}, local work denoted by \(\mathcal{LW}\) \cite{horodecki2002},  total correlations quantified via mutual information \cite{henderson2001,Winter_PRA_2005}, quantum correlation measures which include logarithmic negativity \cite{Werner_PRA_2002}, quantum discord \cite{ollivier2001,prabhu2012, Modi_RMP_2012, bera2017}, work deficit \cite{horodecki2002,aditi2003,aditi2005} (see Appendix \ref{cc_measures} for definitions of measures used in this work). Moreover, in case of CCors, $\chi$ can be chosen to be any classical correlators, \(C_{j_1 j_i}, (j_k=x,y,z)\) or one can choose \(\mathcal{D}^{C^{\max}}\). Note that in case of quantum correlations, this quantity, $\mathcal{D}^{\chi}$ can have a strict upper bound, \(\chi(\rho_{B_1:B_2\ldots B_N})\) with \(\chi\) being the squared of concurrence \cite{Coffman_PRA_2000}, negativity \cite{Ou2007}, discord \cite{Prabhu_PRA_2012, Giorgi11}, teleportation fidelity \cite{Lee_PRA_2009}, dense coding capacity \cite{Prabhu_PRA_2013}. Although no upper bound exists for classical correlation measures, it was shown that a weak upper bound for CC measures can also be obtained for Haar uniformly generated states \cite{saptarshi2021}. In a similar spirit, we know that CC can remain unaltered or can increase with the variation of noise,  and hence, it will be interesting to explore whether such a behavior is universal for other CC measures and for different noise models.

\section{Set of classical correlators under noise: Initial vs. final}
\label{sec:CCnoise}

When each qubit of an arbitrary multiqubit state is sent through different noisy channels, quantum correlations typically deteriorate with the increase of noise. It is interesting to investigate whether CCors behave differently than QCs in presence of noise. In particular, we will establish a relation between the set of classical correlators of the original states and the resulting noise-affected states, thereby revealing the behavior of CC with noise. 

Let us first present the results by fixing the quantum channels. We then compare their patterns against different noise models and show that the detrimental nature of CCors is capable of identifying the channels under use. The output state obtained after individual qubits are passed through noisy channels can be written as \(\Lambda_{B_1}\bigotimes\ldots\bigotimes\Lambda_{B_N}\rho_{B_1\ldots B_N}\) where the set of Kraus operators, \(\{K^{i}_{\mu}\}\) specifies the particular channels  (See Appendix \ref{channels}), with subscript being the site on which it acts and the superscript being the Kraus operators for a given channel. Hence we have 
\begin{equation}
 \rho_{in}^N\rightarrow\rho_f^N=\Lambda\left(\rho_{in}^N\right)=\sum_{\mu,\nu}\bigotimes_{i=1}^{N}K^{i}_{\mu} \rho_{in}^N {K_{\nu}^{i}}^\dagger,
\end{equation}
where the upper limit in summation of \(\mu,\nu\) are the power \(N\) of the maximum number of Kraus operators.


{\it Effects of dephasing channels. }  We begin by investigating the impact of phase damping channels on arbitrary multipartite states.

\noindent\textbf{Proposition I.} {\it  Any arbitrary CCors of an arbitrary multipartite state, \(\rho_{B_1 \ldots B_N}\) in the \(xy\) plane decays while it remains constant in the \(z\)-direction. }

\begin{proof}
 To perform such analysis, let us consider the action of individual Kraus operator on Pauli matrices, i.e.,  $K_0 \sigma_i K_0^\dagger=\left (1-\frac{p}{2}\right)\sigma_i$ $(i=x,y,z)$ and $K_1 \sigma_i K_1^\dagger=- \left (\frac{p}{2}\right)\sigma_i$ $(i=x,y)$ and $K_1 \sigma_z K_1^\dagger= \left (\frac{p}{2}\right)\sigma_z$. Note also that both the operators are Hermitian, i.e., $K_0=K_0^\dagger$ and  $K_1=K_1^\dagger$. 
After sending all the parties of $\rho_{B_1\ldots B_N}$ through the phase damping channel, having a noise parameter, \(p\), any  arbitrary initial CCors, say, $k$-partite CCors in the \(xy\)-plane transform as 
\begin{eqnarray}
 C^{deph}_{j_{1}\ldots j_{k}} &=& \sum_{r=0}^{N}\left [\sum_{q=0}^{\min\{r,k\}}(-1)^q\binom{k}{q}\binom{N-k}{r-q}\right ] \times \nonumber \\
&&\left(\frac{p}{2}\right )^r \left (1-\frac{p}{2}\right )^{N-r}C_{j_{1}\ldots j_{k}}, \,
 k =1, \ldots, N,
\label{eq:CCordeph}
\end{eqnarray}
   where \(N-k \ge r -q\), otherwise it vanishes. In the above expression, $q$ counts the number of noncommutating Pauli matrices with $K_1$. 
   
   Notice that  when all of them are in the $z$-direction, any correlators remain unaltered after phase damping channel, i.e.,  \(C^{deph}_{z_1 \ldots z_k} = C_{z_1 \ldots z_{k}}\) (\(k=1, \ldots N\))  since \(\sigma_z\) commutes with $K_1$. Therefore, we obtain the connection between \(\mathcal{S}^{CC}\) of  \(\rho_{B_1\ldots B_N}\) and that of resulting state after passing through local phase damping channels, \(\rho_{B_1\ldots B_N}^{deph}\).  
\end{proof}


When all the \(N\) parties are affected by noise, the relation between the initial and the final classical correlators of any two-party reduced states remains the same even when the state does not possess any permutational symmetry. Therefore, we can find the distributed CCors,  \(\mathcal{D}^{C^{deph}} (\rho_{B_1\ldots B_N})\) after the noise in terms of the original CCors, \(\mathcal{D}^{C} (\rho_{B_1\ldots B_N})\), given by 
\begin{eqnarray}
&&\mathcal{D}^{C^{deph}} (\rho_{B_1\ldots B_N}^{deph}) =  (1 - \frac{p}{2})^{N-1} (1 - \frac{5 p}{2} + \frac{Np}{2}) \nonumber \\
&&+ \sum_{r=2}^N \left [\sum_{q=0}^2 (-1)^q\binom{2}{q}\binom{N-2}{r-q}\right ] 
\left(\frac{p}{2}\right )^r \left (1-\frac{p}{2}\right )^{N-r}\nonumber 
\\ && \sum_{i=2}^N C_{j_1 j_{i}},
\label{eq:distriCcorPDC}
\end{eqnarray}
since each noisy CCors, \(C^{deph}_{j_1j_i}\) are connected with noiseless \(C_{j_1j_i}\) in a similar fashion. 

{\it CCors under depolarizing noise.} For phase damping channel, connection between CCors of states before and after the actions of noisy channels was possible since the transformation of Pauli matrices under Kraus operators can be found. In case of depolarising channels, Kraus operators involve all Pauli matrices and hence the calculation becomes tedious due to the increased number of possibilities of commuting algebra. However, we obtain a decay pattern  for genuine $N$-party CCors.

\noindent\textbf{Proposition II.}  {\it A compact relation  of genuine CCors between input-output noisy pairs implies exponential decay with the increase of the strength of the noise.}   

\begin{proof}
 When all the qubits of a \(N\)-party state, \(\rho_{B_1 \ldots B_N}\) are passed through depolarizing channels individually, the genuine CCors 
 takes the form as  
\begin{eqnarray}
    C^{DPC}_{j_1\ldots j_N} &=& \left [\sum_{\underset{r+s+t+u=N}{r,s,t,u}}^{N}(-1)^{t+u}\frac{N!}{r!s!t!u!}\right ] \times \nonumber \\
&&\left(1-p\right )^r \left (\frac{p}{3}\right )^{s}\left (\frac{p}{3}\right )^{t}\left (\frac{p}{3}\right )^{u}C_{j_{1}\ldots j_{N}}\nonumber\\&&=\left(1-\frac{4p}{3}\right)^NC_{j_{1}\ldots j_{N}}, j=x,y,z, 
\label{eq:CCordepo}
\end{eqnarray}
where variable $t$ and $u$ counts the number of noncommuting matrices with $K_i$. Unlike PDC,  all $j_k$'s are taken to be the same Pauli matrix in this case.
\end{proof}

{\it The aftermath of amplitude damping noise on CCors.} The action of ADC on states is quite different compared to the PDC and DPC. In particular, it only acts when the state is in an excited state. To obtain a relation between the initial and the noisy final states, we first notice that \(K_0 \sigma_i K_0^{\dagger} = \sqrt{1-p} \sigma_i\) (\(i=x, y\)) while in case of \(\sigma_z\), 
\(K_0 \sigma_z K_0^\dagger = |0\rangle\langle 0| - (1 - p) |1\rangle \langle 1|\). On the other hand, \(K_1 \sigma_i K_1^{\dagger}\)  vanishes with \(i=x, y\) and \(K_1 \sigma_z K_1^{\dagger} = p |1\rangle \langle 1|\). Hence any classical correlator in the \(xy\)-plane transforms as
\begin{eqnarray}
     C^{ADC}_{j_{1}\ldots j_k} =(1-p)^{k/2} C_{j_{1}\ldots j_k}.
\label{eq:CCoradc}
\end{eqnarray}
In this case, the distributed CCors in the \(xy\)-plane reads
\begin{eqnarray}
&&\mathcal{D}^{C^{ADC}} (\rho_{B_1\ldots B_N}^{ADC}) = (1-p)  \sum_{i=2}^N C_{j_1 j_{i}}.
\label{eq:distriCCorADC}
\end{eqnarray}
Introducing $\sigma_z$ in any site of the CCors makes the calculation more involved. To illustrate the CCors in the \(z\) direction, let us consider the generalized \(GHZ\) state (gGHZ), \(|gGHZ^N \rangle = \cos \frac{\theta}{2} |0^{\otimes N}\rangle + \exp(i \phi) \sin \frac{\theta}{2} |1^{\otimes N}\rangle\) where any reduced  two-party CCors are \(C_{xx} = C_{yy} =0\) and \(C_{zz} =1\).  After sending all the parties through noisy ADC, classical correlators in the \(xy\)-plane vanish as also seen from Eq. (\ref{eq:CCoradc}) while \(C^{ADC}_{zz} = 1 - 2 p (1 - p)  (1-  \cos \theta)\) which leads to \(\mathcal{D}^{C^{ADC}_{zz}} (|gGHZ^N\rangle^{ADC}) = (N-1) C^{ADC}_{zz} \). 

%
Among the set of  three-qubit states, there exist two distinct  SLOCC inequivalent classes, namely the \(GHZ\)- and the \(W\)-class states  \cite{Dur_PRA_2000}. Here we present $k$-party correlator for a \(N\)-qubit generalized \(W\) state, represented by \(\ket{gW^N}=\sum_{i=1}^{N}a_i\mathcal{P}(\ket{00\ldots1})\) \cite{aditi2003}  where $\mathcal{P}$  denotes all the permutations and $\sum_{i=1}^{N}\abs{a_i}^2=1$. We calculate $k$-party non-genuine as well as genuine classical correlators if the state is passed through ADC. Instead of finding the relation between input-output CCors, we calculate the output state obtained after local ADC acts on each qubit. First we notice that \(K_0\ket{0}=\ket{0}\), \(K_0\ket{1}=\sqrt{1-p}\ket{1}\), \(K_1\ket{0}=0\) and \(K_1\ket{1}=\sqrt{p}\ket{0}\). After application of local channels, the final state is given by
    \begin{equation}
        \rho_f=(1-p)\ket{gW^N}\bra{gW^N}+p\ket{0\ldots 0}\bra{0\ldots0}.
        \label{eq:gw}
    \end{equation}
Hence $k$-party correlator in the $z$ direction is \(C^{ADC}_{z_1z_2\ldots z_k}=p+(1-p)\left[\sum_{i=1}^{N}(-1)^{\theta(k-i)}\abs{a_i}^2\right]\), where \(\theta(x)=1\) for \(x\le 0\) and \(\theta(x)=0\) for \(x> 0\). Any classical correlators, both genuine and non-genuine, in the \(xy\)-plane vanish except two-party CCors due to the inherent symmetry in the two-party reduced state of the $|gW^N\rangle$ state. Hence, \(C_{xx}=C_{yy}=(1-p)(a_1a_i^{*}+a_ia_1^{*})\) and also two-body correlator in the $z$-direction is \(C_{zz}^{1i}=p+(1-p)(1-2\abs{a_1}^2-2\abs{a_i}^2)\), \(i=2, \ldots, N\).


\subsection{Discrimination of channels via classical correlators }
\label{subsec:discrimination}

Suppose the initial  state shared between parties situated in distant locations is sent through local noisy channels and some properties of the resulting state is  known. However, the channels acting on qubits are unknown. It is a typical scenario in quantum key distribution \cite{Scarani_RMP_2009,Feihu_RMP_2020} and in many other  quantum communication protocols. There are procedures based on entanglement measure \cite{Piani_PRL_2009}, discord monogamy score by which one can discriminate quantum channels provided  the set is known. When the set consists of PDC, ADC, and DPC, we will demonstrate that the behavior of classical correlators which can be measured easily in laboratories  can be used to discriminate them. 

\begin{figure}
		\includegraphics[width=\linewidth]{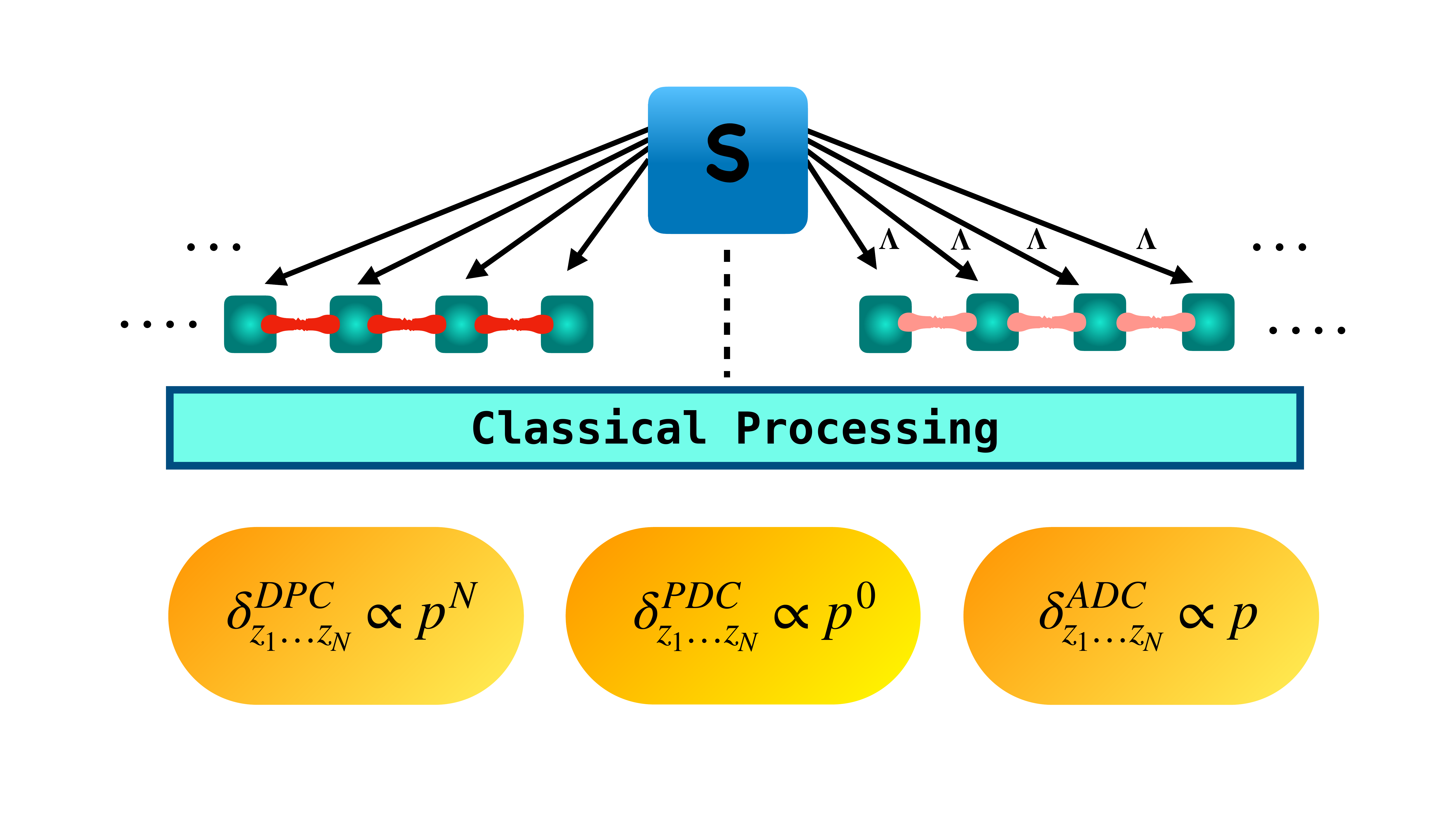}
		\caption{\textbf{A schematic diagram for discrimination of channels.} We provide a general sketch to discriminate three channels, PD, AD and DP channels which act locally on states. 
		A source, \(S\), produces two copies of a \(N\)-party state -- CCors of the first copy is calculated when each qubit is sent through noiseless channels while CCors of the second copy of the state is affected by local noisy channels, \(\Lambda\). In the step of classical processing, we calculate difference between genuine CCors in the \(z\)-direction before and after application of quantum channels, where \(\ket{gW^N}\) is produced as an initial state. A nonlinear variation with noise strength depicts the channel to be depolarising, linear one represents amplitude damping channel while when the difference vanishes,  the channel is phase damping.}
		    \label{fig:cc_schematics}
\end{figure}

In this process, a class of three-qubit entangled state, specifically, \(\ket{gW^3} =\cos\alpha\ket{001}+e^{i\gamma_1}\sin\alpha\cos\beta\ket{010}+e^{i\gamma_2}\sin\alpha\sin\beta\ket{100}\) can be taken as resource. A schematic diagram for the scheme is depicted in Fig. \ref{fig:cc_schematics}. Initially, a source  produces a $\ket{gW^3}$ state and sends it  
to $B_1$, $B_2$ and $B_3$, who measure CCors. In the second situation,   noise acts uniformly on each qubit of the state prepared by the source during sending and again they measure CCors.

\begin{figure*}
		\includegraphics[width=\linewidth]{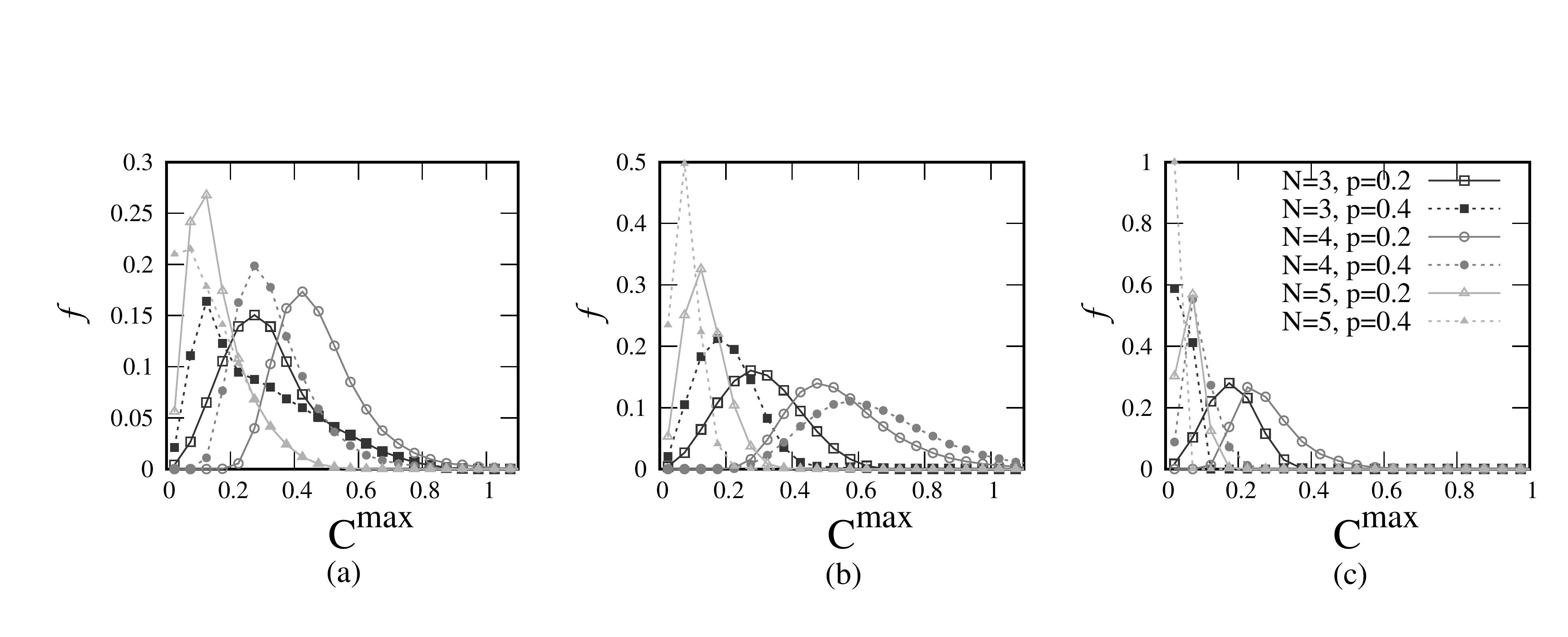}
		\caption{\textbf{Plot of normalized frequency distribution (NFD) (ordinate) of maximum genuine classical correlators, \(C^{\max}\) (abscissa).} NFD is computed for $10^5$ Haar uniformly generated states for different noise strength and for increasing number of parties (squares, circles and triangles are for \(N=3, 4\) and \(5\) respectively). Panel (a), (b) and (c) describe the NFD for phase damping, amplitude damping and depolarising channels respectively. Decreasing the shades of the plot indicate the increasing number of parties and solid lines represent the noise strength,  \(p=0.2\) in the channel and   the  dashed line indicates \(p=0.4\). All axes are dimensionless.}
		    \label{fig:gen_cmax}
\end{figure*}

\begin{enumerate}

    \item First notice that \(C_{zz} (\rho_{B_1 B_2}^{deph}) = 1 \) which leads to \(\mathcal{D}^{C^{(deph)}_{zz}} =2\), i.e., it is constant, independent of the state and noise parameters of the PDC. In case of depolarizing channel, \(\mathcal{D}^{C^{(DPC)}_{zz}} = 2(1- \frac{4p}{3})^2 e^{2i\gamma_2}\sin^2\alpha\sin^2\beta \), thereby suppressing CCors quadratically while \(\mathcal{D}^{C^{ADC}_{zz}} =2 \left(\sin ^2\alpha  \left((2 p-1) \sin ^2\beta +p \cos ^2\beta \right)+p \cos ^2\alpha \right)\). Both of them depend on state and noise parameters although the later decreases with noise linearly. Clearly, the detrimental effects on \(\mathcal{D}^{C_{zz}}\) with the increase of \(p\) is much more drastic in case of depolarizing channel while it is one order of magnitude less for ADC, and CCors remain unaffected by phase damping channel. If the state does not change, we conclude that the channel is dephasing while if the CCor of the resulting state decreases with noise linearly, it is ADC and otherwise, depolarizing channel.  
    
    \item Like distributed CCors, similar dependence on noise parameter of genuine classical correlators can also identify the channels which act on all the qubits. If one considers \(C_{zzz}\) with the input state being the \(gW\) state, it again remains unaltered under  dephasing channel, behaves linearly with the noise parameter in ADC  as \(C_{zzz}^{ADC} = -(1-2p)\). When the depolarizing channel acts on all the qubits, the state depends on cubic power of  the noise parameters, i.e., \(C_{zzz}^{DPC}=(1-\frac{4p}{3})^3 \left(\cos^2\alpha+\sin^2\alpha\left(e^{2i\gamma_1}\cos^2\beta+e^{2i\gamma_2}\sin^2\beta\right)\right) \). 
    
    \item Instead of three-qubit initial state, if an $N$-qubit \(\ket{gW^N}\) state is shared, we can again differentiate the channels by looking into the \(N\)-party CCors in the $z$ direction. It remains unaltered if the state passes through the dephasing channel, damped linearly with noise parameter as depicted by Eq. (\ref{eq:gw}) for ADC while for DPC, it changes as $C_{z_1z_2\ldots z_N}^{DPC}=\left (1-\frac{4p}{3}\right)^NC_{z_1z_2\ldots z_N}$. For DPC $N$-party genuine  CCors get affected nonlinearly with noise and depends upon the system size. Hence, by measuring CCors we can distinguish the channels.
    
\end{enumerate}

\textbf{Remark 1.} It is interesting to notice that distributed CCors in the \(xy\)-plane are not always capable to distinguish all the channels. 

\textbf{Remark 2.} Instead of the \(gW\) state, if the shared state is the generalized \(GHZ\) state, the distributed CCors in the \(z\) direction does not work as efficiently as shown for the shared \(gW\) state. If \(\mathcal{D}^{C_{zz}}\) is constant, we can immediately infer that the channel is dephasing. However, \(\mathcal{D}^{C^{DPC}_{zz}} = 2 (1 - \frac{4p}{3})^2\) and \(\mathcal{D}^{C^{ADC}_{zz}}=2-4p(1-p)(1-\cos\theta)\)  which imply that they both decrease quadratically with \(p\) although the former does not depend on the state parameters while the latter depends. 


Similarly, in case of shared \(|gGHZ\rangle\), \(C_{zzz}\) can also serve the purpose with the help of \(C_{xxx}\). Again the constant value with the increase of noise implies the dephasing channel. For a fixed state, dependence on \(p\) of \(C_{zzz}^{ADC (DPC)}\) is same and hence they cannot distinguish these two channels.  However, \(C_{xxx}^{ADC} = (1 - p)^{3/2} \cos \phi \sin \theta\) which varies with the power of \(p^{3/2}\) while  \(C_{xxx}^{DPC} = (1 - \frac{4p}{3})^3 \cos \phi \sin \theta \sim \mathcal{O}(p^3)\), thereby discriminating the channels. 

\begin{figure*}
		\includegraphics[width=20cm, height=15cm]{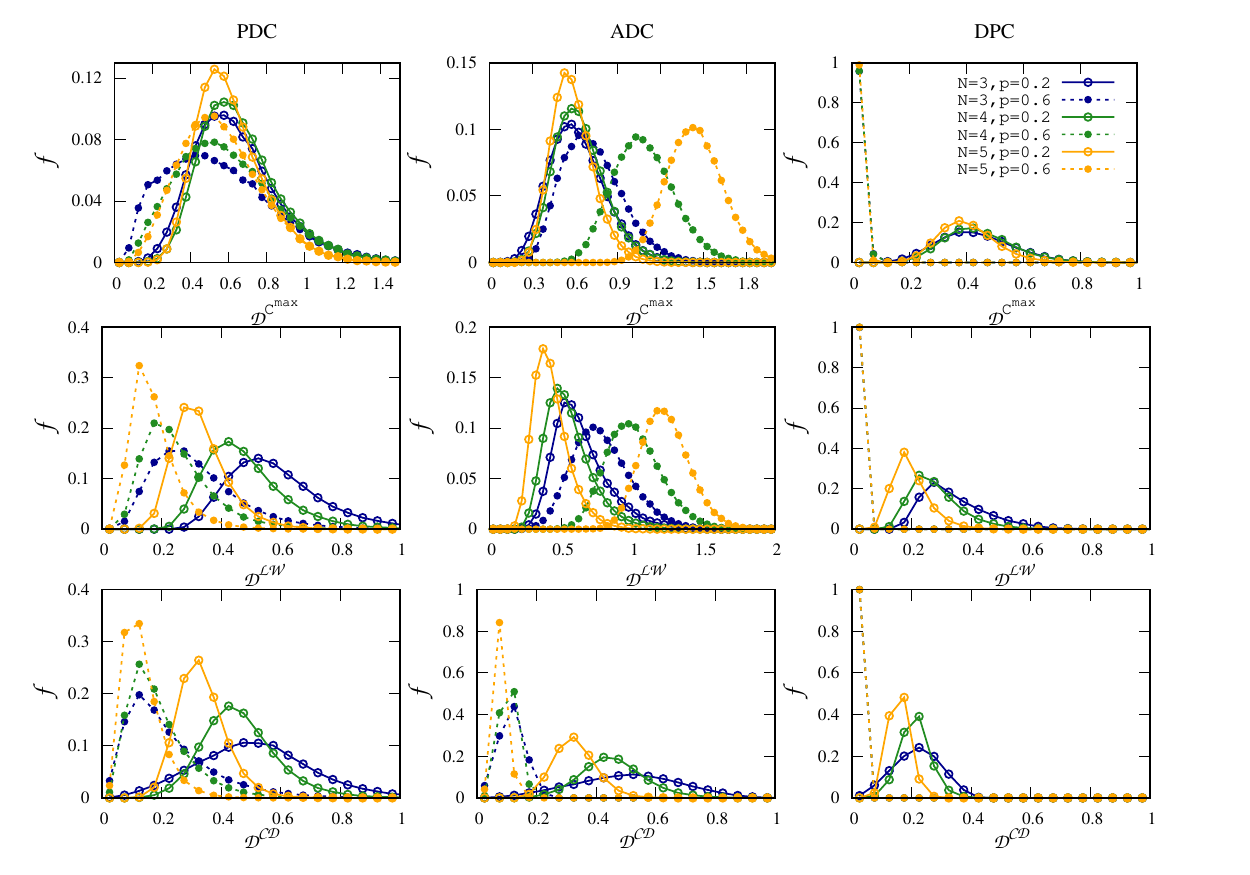}
		\caption{Normalised frequency distribution (vertical axis) for Haar uniformly generated three-, four- and five- qubit states. Horizontal axis of first, second and third row are respectively \(\mathcal{D^{C^{\max}}}\), \(\mathcal{D^{LW}}\) and \(\mathcal{D^{CD}}\). When PDC (first column), ADC (second column) and DPC (third column) act on each qubit, the respective CC measures are calculated for a fixed noise strength, \(p\). With the increase of \(N\), $f$ shifts towards left. With the increase of noise, distribution again shifts toward left except some scenarios with ADC. All the axes are dimensionless.}
		    \label{fig:NFD_345}
\end{figure*}

\section{Universal decay nature by Classical correlations against noise}
\label{sec:frequency}

In contrast to entanglement, we have already shown that CCors in some fixed directions may not change with the variation of noise. Let us study how the shareability of  CC measures including the axiomatic ones behave when each qubit of Haar uniformly generated states are passed through noisy channels. In other words, we want to find whether CC measures of multipartite state are less fragile than the QC measures or not and  proceed by looking into the changes that occurred in the frequency distribution of classical correlation measures after an individual qubit of a multipartite state passes through noisy channels. 

To quantify the trends  of the normalised frequency distribution, we calculate the moments of distributed CC measures, i.e., we compute mean (\(\langle \mathcal{D}^\chi \rangle\)), standard deviation ($\sigma_{\mathcal{D}^\chi}(p)$), and skewness ($\eta_{\mathcal{D}^\chi}(p)$) of the distribution for a fixed CC measure obtained from the randomly generated states with a fixed noise strength of a given channel. 

{\it Generation of random states.}  A generic $N$-party pure state $\ket{\psi}$ in the computational basis can be represented as  $\ket{\psi} = \sum_{l=0}^{2^{N}-1} a_{l} \ket{i_{1}i_{2}...i_{N}} \quad \textrm{where} \quad \sum_{l=0}^{2^{N}-1} \lvert a_l \lvert ^{2} =1 $. For Haar uniform generation,  the coefficients $a_l$s are taken   to be  complex  and hence we can write them as $a_l = \alpha_l + i \beta_l $ where $\alpha_l$ and $\beta_l$ are real numbers and the values are generated from a Gaussian distribution of vanishing mean and unit standard deviation \cite{bengtsson2006}. By generating \(10^{5}\) random states which are sent through different noisy channels, we compute the genuine classical correlators and distributed CC measures based on CCors, axiomatic CC measures such as classical correlation measure associated with quantum discord, local work \cite{horodecki2002}. Since there exist examples of  states and channels for which they remain constant as mentioned before, like QCs, the decaying behavior of CC measures with the increment  of noise for generic  states is not apriori guaranteed. 


{\it Trends of maximal and distributed genuine  CCors.} As discussed in Sec. \ref{sec:CCnoise}, genuine CCors decay with the increase of noise, irrespective of the channels. Specifically, from the normalized frequency distribution (NFD) of $C^{\max}$, which is defined as
\begin{equation}
   f(C^{\max})=\frac{\text{Number of states with fixed}\quad C^{\max}}{\text{Total number of states generated}},
   \label{eq:nfd}
\end{equation}
we find that for a fixed number of parties, $N$, 
the mean of the distribution, $\langle C^{\max}\rangle$, clearly decreases with the increase of noise as shown in Fig. \ref{fig:gen_cmax}. Moreover, we find that in all the channels considered, for a fixed noise strength in the channel, average $C^{\max}$ remains almost constant (slightly increases) when one increases the number of parties from three to four while it decreases when number of parties increases from four to five. We know that with the increase of number of qubits, the mean entanglement in multipartite states generated Haar uniformly increases, and the results show that CCors in highly entangled states are more fragile than that of the weakly entangled states. Let us investigate whether all CC measures behave similarly or not.

{\it Behavior of distributed CC measures.} We Haar uniformly generate $10^5$ three-, four- and five-qubit states and find the behavior of  distributed CC measures, \(\mathcal{D}^\chi\) where $\chi$ is either ${C^{\max}}$, $\mathcal{CD}$ and $\mathcal{LW}$ by varying the strength of noise. Analyzing frequency distributions for CCors, both $C^{\max}$ and \(\mathcal{D}^{C^{\max}}\), $\mathcal{D}^\mathcal{CD}$ and $\mathcal{D}^\mathcal{LW}$, we find that the behavior of CCors are in some situation differ from the axiomatic CC measures. Specifically, we observe the following:

\begin{enumerate}
    \item {\it Decay with noise.} Both CCors \(\mathcal{D}^{C^{\max}}\) and \(C^{\max}\)  and axiomatic CC measures decay with the increase of $p$ for any number of parties.  Notice that such decaying behavior can also be seen for QC measures.
    
    \item {\it Patterns of NFD.}  NFD of any CC measures, \(f (\mathcal{D}^{C^{\chi}})\), as in Eq. (\ref{eq:nfd}) take Bell-like shape in presence of noise except in case of high depolarizing noise as depicted in Fig. 
    \ref{fig:NFD_345}. Depending on the channel, the maxima and the spread of the distribution change.  The average value of the frequency distribution also decreases with \(p\). It is independent of the channels applied on each qubit.
    
    \item {\it Behavior of average CC with number of parties.} For a fixed noise \(p\), axiomatic CC measures behave differently with the increase of \(N\) than the CCors. In particular, \(\langle \mathcal{D}^{C^{\max}}\rangle\) increases when \(N\) goes from three to four while it decreases when \(N\) goes beyond four. On the other hand, for axiomatic measures, it is not the case, i.e., \(\langle \mathcal{D}^{\mathcal{CD}}\rangle\) and \(\langle \mathcal{D}^{\mathcal{LW}}\rangle\) decreases with \(N\) (see Tables \ref{tab:mean_noise}, \ref{tab:mean_noise_lw} and \ref{tab:mean_noise_CD} ).
    
    \item {\it Spread of CC.} The standard deviation $(\sigma_{\mathcal{D^{\chi}}}= \sqrt{ \langle \mathcal{D}^{\mathcal{X}^2}\rangle - \langle \mathcal{D}^{\mathcal{X}}\rangle^2}) )$  of the CC measures typically decreases for Pauli channels with the increase of $p$  except $\mathcal{X}=C^{\max}$ or except genuine CCors while for ADC, it remains constant or fluctuates for a fixed \(N\). 
    
    \item {\it Broken symmetry.} We also compute skewness \(\left (3 \cross \frac{\langle\mathcal{D^{\chi}}\rangle-\Bar{\mathcal{D^{\chi}}}}{\sigma_{\mathcal{D^{\chi}}}}\right) \), where \(\Bar{\mathcal{D^{\chi}}}\) is median, for each distributions with a fixed channel and $N$ to evaluate how symmetry of the distribution gets effected with noise and with number of qubits. It shows that the symmetry around mean CCs is not present in multipartite states. With the variation of noise or number of parties, no universal pattern emerges in this case.
\end{enumerate}

The entire analysis reveals that the trends of distributed axiomatic CC measures with noise for Haar uniformly generated states resemblance more with the decay of pattern of QC  compared to the genuine as well as distributed CCors which show some anomaly. Moreover, the decaying nature of CC measures becomes prominent with noise for depolarising channel while the effects of PDC and ADC are weak on CCs. The patterns of frequency distributions, average CCs and SDs support such hierarchy among channels. Moreover, our study shows that although CC measures have definite mean, the decrease of average distributed CC measures is not additive with the increase of number of qubits.

\small 
\begin{table}
\begin{tabular}{c}
Phase-flip channel\\
\scalebox{0.8}{
\begin{tabular}{|c|c|c|}
\hline
$p=0.2$ & $p=0.4$ & $p=0.6$  \\
\hline
          \begin{tabular}{c|c|c|c}
            $N$  & $3$ & $4$  & $5$  \\ 
          \hline 
          $\langle \mathcal{D^{{C}^\text{max}}} \rangle (p)$   &$0.639$  &  $0.661$ & $0.609$  \\
          \hline
         $\sigma_\mathcal{D^{{C}^\text{max}}} (p)$   &$0.225$  &$0.212$ & $0.179$  \\
         \hline
         $\eta_\mathcal{D^{{C}^\text{max}}}(p)$   & $-2.968$ &$-1.026$ & $0.284$  \\
         \end{tabular}         
         &         
          \begin{tabular}{c|c|c}
          $3$ & $4$ & $5$ \\
          \hline 
          $0.567$  &   $0.600$ & $0.564$\\
          \hline
         $0.260$  & $0.245$ & $0.207$  \\
         \hline
         $-2.279$ & $0.734$ & $0.376$ \\
         \end{tabular}             
         &         
          \begin{tabular}{c|c|c}
          $3$ & $4$ & $5$ \\
          \hline 
          $0.549$ & $0.590$ & $0.559$  \\
          \hline
          $0.279$  &$0.257$ & $0.214$  \\
         \hline
         $1.394$ & $-1.681$ & $0.294$ \\
         \end{tabular}  \\       
\hline          
\end{tabular}}\\
Depolarizing Channel  \\
\scalebox{0.75}{
\begin{tabular}{|c|c|c|}
\hline 
$p=0.2$ & $p=0.4$ & $p=0.6$  \\
\hline 
          \begin{tabular}{c|c|c|c}
          $N$ & 3 & 4 & 5 \\ 
          \hline 
          $\langle \mathcal{D^{{C}^\text{max}}} \rangle$ (p) & $0.432$   & $0.443$ & $0.404$  \\
          \hline
         $\sigma_\mathcal{D^{{C}^\text{max}}}(p)$ & $0.129$  & $0.119$ & $0.099$  \\
         \hline
         $\eta_\mathcal{D^{{C}^\text{max}}}(p)$ & $-0.286$   & $-0.611$ & $0.272$  \\
         \end{tabular}         
         &         
          \begin{tabular}{c|c|c}
          3 & 4 & 5 \\ 
          \hline 
           $0.174$  & $0.179$  &   $0.163$\\
          \hline
          $0.052$  &$0.048$  &$0.040$   \\
         \hline
          $1.444$  & $-0.610$ & $0.225$  \\
         \end{tabular}             
         &         
          \begin{tabular}{c|c|c}
          3 & 4 & 5 \\ 
          \hline 
          $0.032$  & $0.032$ & $0.030$  \\
          \hline
         $0.009$ & $0.008$ &$0.007$   \\
         \hline
          $-1.508$  & $-0.611$ & $0.428$  \\
         \end{tabular}      \\   
\hline                      
\end{tabular}}\\
Amplitude-damping Channel\\
\scalebox{0.80}{
\begin{tabular}{|c|c|c|} 
\hline 
$p=0.2$ & $p=0.4$ & $p=0.6$\\
\hline 
          \begin{tabular}{c|c|c|c}
          $N$ & 3 & 4 & 5 \\ 
          \hline 
          $\langle \mathcal{D^{{C}^\text{max}}} \rangle$ (p) & $0.616$   & $0.634$ &$0.584$   \\
          \hline
         $\sigma_\mathcal{D^{{C}^\text{max}}}(p)$ & $0.193$  &$0.179$  & $0.148$  \\
         \hline
         $\eta_\mathcal{D^{{C}^\text{max}}}(p)$ &$1.291$  & $-1.022$ &$-0.101$   \\
         \end{tabular}         
         &         
          \begin{tabular}{c|c|c}
          3 & 4 & 5 \\ 
          \hline 
         $0.507$  & $0.579$  & $0.664$  \\
          \hline
        $0.182$    & $0.190$ & $0.201$  \\
         \hline
        $-2.532$   & $0.915$ &  $0.298$ \\
         \end{tabular}             
         &         
          \begin{tabular}{c|c|c}
         3 & 4 & 5 \\ 
          \hline 
        $0.720$    & $1.079$ & $1.440$  \\
          \hline
         $0.209$    &$0.212$  &$0.195$   \\
         \hline
        $0.090$  &$2.120$  & $0.092$  \\
         \end{tabular}         \\
\hline                                           
\end{tabular}}\\

\end{tabular}
\caption{Statistical quantities  of the normalized frequency distributions of $\mathcal{D^{{C}^\text{max}}}$ for different strength of noise and channels.  $\langle \mathcal{D^{{C}^\text{max}}} \rangle (p)$,  $\sigma_\mathcal{D^{{C}^\text{max}}}(p)$ and $\eta_\mathcal{D^{{C}^\text{max}}}(p)$  denote the average, standard deviation and skewness of the distribution respectively.  The data analysis contains \(10^5\) Haar uniformly generated states.  }
\label{tab:mean_noise}
\end{table}
\normalsize

\small
\begin{table}
\begin{tabular}{c}
Phase-flip channel\\
\scalebox{0.80}{
\begin{tabular}{|c|c|c|}
\hline
$p=0.2$ & $p=0.4$ & $p=0.6$  \\
\hline
          \begin{tabular}{c|c|c|c}
            $N$  & $3$ & $4$ & $5$ \\ 
          \hline 
          $\langle \mathcal{D^{LW}} \rangle$ (p)   &$0.583$  &  $0.478$ & $0.331$ \\
          \hline
         $\sigma_\mathcal{D^{LW}}(p)$   &$0.160$  &$0.132$ & $0.091$   \\
         \hline
         $\eta_\mathcal{D^{LW}}(p)$   & $-1.641$ &$0.699$ & $4.021$ \\
         \end{tabular}         
         &         
          \begin{tabular}{c|c|c}
          $3$ & $4$ & $5$ \\
          \hline 
          $0.415$  &   $0.335$ & $0.231$ \\
          \hline
         $0.149$  & $0.121$ & $0.081$  \\
         \hline
         $-6.687$ & $0.041$ & $1.333$  \\
         \end{tabular}             
         &         
          \begin{tabular}{c|c|c}
          $3$ & $4$ & $5$ \\
          \hline 
          $0.316$ & $0.252$ & $0.173$  \\
          \hline
          $0.146$  &$0.116$ & $0.077$ \\
         \hline
        $1.593$ & $-0.410$ & $0.818$  \\
         \end{tabular}  \\       
\hline          
\end{tabular}}\\
Depolarizing Channel  \\
\scalebox{0.80}{
\begin{tabular}{|c|c|c|}
\hline 
$p=0.2$ & $p=0.4$ & $p=0.6$  \\
\hline 
          \begin{tabular}{c|c|c|c}
          $N$ & 3 & 4 & 5 \\ 
          \hline 
          $\langle \mathcal{D^{LW}} \rangle$ (p) & $0.341$   & $0.282$ & $0.197$  \\
          \hline
         $\sigma_\mathcal{D^{LW}}(p)$ & $0.109$  & $0.087$ & $0.059$  \\
         \hline
         $\eta_\mathcal{D^{LW}}(p)$ & $0.938$   & $1.243$ & $3.457$  \\
         \end{tabular}         
         &         
          \begin{tabular}{c|c|c}
          3 & 4 & 5 \\ 
          \hline 
           $0.115$  & $0.093$  &   $0.064$\\
          \hline
         $0.048$  &$0.037$  &$0.025$   \\
         \hline
         $-1.121$  & $1.214$ & $-2.160$  \\
         \end{tabular}             
         &         
          \begin{tabular}{c|c|c}
          3 & 4 & 5 \\ 
          \hline 
          $0.019$  & $0.015$ & $0.010$  \\
          \hline
        $0.009$ & $0.007$ &$0.004$   \\
         \hline
        $0.971$  & $1.270$ & $-3.000$  \\
         \end{tabular}      \\   
\hline                      
\end{tabular}}\\
Amplitude-damping Channel\\
\scalebox{0.80}{
\begin{tabular}{|c|c|c|} 
\hline 
$p=0.2$ & $p=0.4$ & $p=0.6$\\
\hline 
          \begin{tabular}{c|c|c|c}
          $N$ & 3 & 4 & 5 \\ 
          \hline 
          $\langle \mathcal{D^{LW}} \rangle$ (p) & $0.641$  & $0.555$ &$0.433$   \\
          \hline
         $\sigma_\mathcal{D^{LW}}(p)$ & $0.199$  &$0.164$  & $0.127$  \\
         \hline
         $\eta_\mathcal{D^{LW}}(p)$ & $1.104$ & $1.408$ &$0.874$   \\
         \end{tabular}         
         &         
          \begin{tabular}{c|c|c}
          3 & 4 & 5 \\ 
          \hline 
         $0.612$  & $0.646$  & $0.667$  \\
          \hline
        $0.219$  & $0.193$ & $0.165$  \\
         \hline
        $-2.285$  & $2.238$ &  $1.036$ \\
         \end{tabular}             
         &         
          \begin{tabular}{c|c|c}
         3 & 4 & 5 \\ 
          \hline 
        $0.784$  & $1.010$ & $1.231$  \\
          \hline
         $0.207$ &$0.193$  &$0.169$   \\
         \hline
        $0.143$  &$2.393$  & $-0.923$  \\
         \end{tabular}         \\
\hline                                           
\end{tabular}}\\

\end{tabular}
\caption{ Statistical quantities for $\mathcal{D^{LW}}$. Other specifications are same as in Table \ref{tab:mean_noise}.}
\label{tab:mean_noise_lw}
\end{table}
\normalsize


\small
\begin{table}
\begin{tabular}{c}
Phase-flip Channel\\
\scalebox{0.75}{
\begin{tabular}{|c|c|c|}
\hline
$p=0.2$ & $p=0.4$ & $p=0.6$  \\
\hline
          \begin{tabular}{c|c|c|c}
            $N$ & $3$ & $4$ & $5$ \\ 
          \hline 
          $\langle \mathcal{\mathcal{D}^{CD}} \rangle$ (p)   &$0.514$  &  $0.464$ & $0.334$ \\
          \hline
         $\sigma_\mathcal{\mathcal{D}^{CD}}(p)$   &$0.182$  &$0.126$ & $0.079$  \\
         \hline
         $\eta_\mathcal{\mathcal{D}^{CD}}(p)$  & $-3.630$ &$-1.830$ & $1.936$   \\
         \end{tabular}         
         &         
          \begin{tabular}{c|c|c}
           $3$ & $4$ & $5$\\
          \hline 
           $0.322$  &   $0.283$ & $0.203$ \\
          \hline
         $0.154$  & $0.109$ & $0.067$  \\
         \hline
          $1.554$ & $-1.763$ & $-0.358$ \\
         \end{tabular}             
         &         
          \begin{tabular}{c|c|c}
           $3$ & $4$ & $5$ \\
          \hline 
           $0.222$ & $0.189$ & $0.134$  \\
          \hline
         $0.144$  &$0.102$ & $0.063$  \\
         \hline
         $0.766$ & $-1.304$ & $-0.285$ \\
         \end{tabular}  \\       
\hline          
\end{tabular}}\\
Depolarizing Channel  \\
\scalebox{0.75}{
\begin{tabular}{|c|c|c|}
\hline 
$p=0.2$ & $p=0.4$ & $p=0.6$  \\
\hline 
          \begin{tabular}{c|c|c|c}
          $N$ & 3 & 4 & 5 \\ 
          \hline 
          $\langle \mathcal{\mathcal{D}^{CD}} \rangle$ (p) & $0.218$   & $0.211$ & $0.158$  \\
          \hline
         $\sigma_\mathcal{\mathcal{D}^{CD}}(p)$ & $0.074$  & $0.049$ & $0.032$  \\
         \hline
         $\eta_\mathcal{\mathcal{D}^{CD}}(p)$ & $-1.801$   & $-1.855$ & $1.218$  \\
         \end{tabular}         
         &         
          \begin{tabular}{c|c|c}
          3 & 4 & 5 \\ 
          \hline 
           $0.032$  & $0.032$  &   $0.025$\\
          \hline
          $0.012$  &$0.007$  &$0.005$   \\
         \hline
         $-0.040$  & $-1.925$ & $3.600$  \\
         \end{tabular}             
         &         
          \begin{tabular}{c|c|c}
           3 & 4 & 5 \\ 
          \hline 
           $0.001$  & $0.001$ & $0.0008$  \\
          \hline
        $0.0004$ & $0.0002$ &$0.0001$   \\
         \hline
         $-0.026$  & $-1.967$ & $6.000$  \\
         \end{tabular}      \\   
\hline                      
\end{tabular}}\\
Amplitude-damping Channel\\
\scalebox{0.8}{
\begin{tabular}{|c|c|c|} 
\hline 
$p=0.2$ & $p=0.4$ & $p=0.6$\\
\hline 
          \begin{tabular}{c|c|c|c}
          $N$ & 3 & 4 & 5 \\ 
          \hline 
          $\langle \mathcal{\mathcal{D}^{CD}} \rangle$ (p) &  $0.504$ & $0.457$ &$0.329$   \\
          \hline
         $\sigma_\mathcal{\mathcal{D}^{CD}}(p)$ & $0.160$  &$0.109$  & $0.070$  \\
         \hline
         $\eta_\mathcal{\mathcal{D}^{CD}}(p)$ &$-6.365$ & $-1.568$ &$-0.214$   \\
         \end{tabular}         
         &         
          \begin{tabular}{c|c|c}
          3 & 4 & 5 \\ 
          \hline 
           $0.261$  & $0.240$  & $0.174$  \\
          \hline
       $0.088$  & $0.060$ & $0.038$  \\
         \hline
         $0.758$  & $-1.55$ &  $0.868$ \\
         \end{tabular}             
         &         
          \begin{tabular}{c|c|c}
           3 & 4 & 5 \\ 
          \hline 
         $0.115$  & $0.107$ & $0.078$  \\
          \hline
           $0.040$ &$0.028$  &$0.018$   \\
         \hline
         $0.577$ &$-1.500$  & $2.833$  \\
         \end{tabular}         \\
\hline                                           
\end{tabular}}\\

\end{tabular}
\caption{Statistical quantities  of the normalized frequency distributions of $\mathcal{\mathcal{D}^{CD}}$. Other specifications are same as in Table \ref{tab:mean_noise}.  }
\label{tab:mean_noise_CD}
\end{table}
\normalsize

\subsection{Patterns of CC measures in random $GHZ$- and $W$-class states}

\begin{figure*}
		\includegraphics[width=20cm, height=15cm]{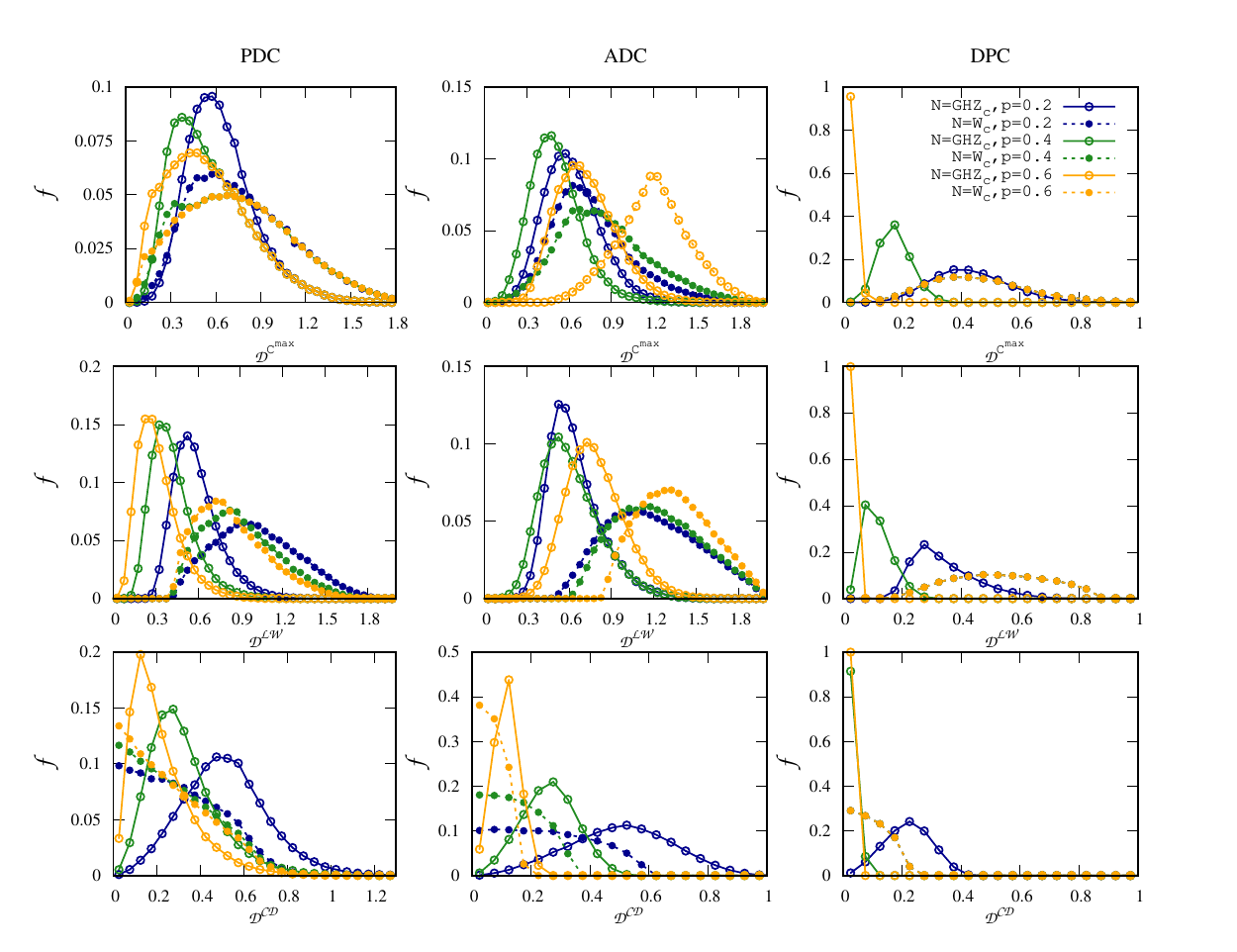}
	\caption{\textbf{CC in \(GHZ\)- vs. \(W\)-class states.} Normalized frequency distribution, $f$ (ordinate) vs. \(\mathcal{D^{\chi}}\) (abscissa), \(\chi\in\{C^{\max} \, (\text{first row}), \mathcal{LW} \, (\text{second row}), \mathcal{CD}\,  (\text{third row})\}\) for three-qubit pure states as inputs. \(W\)-class state are more robust against noise  than the states from the \(GHZ\)-class in terms of distributed CC measures. All other specifications are same as in Fig. \ref{fig:gen_cmax}. All axes are dimensionless.}
		    \label{fig:GHZ-W-lass-NFD}
\end{figure*}

Let us now concentrate on three-qubit pure states as inputs which are disturbed by noise. Since the three-qubit pure states contains two SLOCC inequivalent classes \cite{Dur_PRA_2000}, having distinct entanglement properties, we want to examine whether their CCs also behave differently in presence of noise. In other words, it is intriguing to see if patterns of classical correlation measures produced for states in the $GHZ$-class, (\(\ket{GHZ_c}\)), under noisy channels might disclose certain contrasting traits that cannot be detected for states in the $W$-class (\(\ket{W_c}\)). Haar uniform generation of three-qubit states always lead to $GHZ$-class state while to simulate $W$-class state, one considers
 \(   \ket{{W_c}} = a_{0} \ket{000} + a_{1} \ket{001} + a_{2} \ket{010} + a_{3} \ket{100}\),  
where $\sum_{i}^{3} \lvert a_{i} \rvert ^{2} = 1 $ and we choose each $a_i$ from a Gaussian distribution of mean zero and standard deviation unity \cite{ratul2020}. When each qubit of the   random states from both the classes are sent through noisy channels, the normalized frequency distribution of $\mathcal{D^\chi}$ where $\chi\in\{C^{\max},\mathcal{LW},\mathcal{CD}\}$ for PD, DP, AD channels are depicted in Fig. \ref{fig:GHZ-W-lass-NFD}. We observe that both \(\langle\mathcal{D}^\chi\rangle\) and \(\sigma_{\mathcal{D}}\) for states from the $GHZ$- and the $W$-class possesses similar feature with the increase of noise parameter \(p\) for all the channels except ADC. Among three-qubit states, we find that average distributed CCs for the $W$-class states which typically possesses less genuine multipartite entanglement are more robust against noise than that of the $GHZ$-class  states  (see Table \ref{tab:GHZ-W-NFD-D}), i.e., $\ev{\mathcal{D}_{W_c}^{\chi}}> \ev{\mathcal{D}_{GHZ_c}^{\chi}}$. 
To make the comparison more concrete,  we define a quantity which exhibits the average decay rate of distributed CC measures with \(p\), defined as
\begin{equation}    \mathcal{R}^{C^{\max}}=\expval{\frac{\expval{\mathcal{D^{C^{\max}}}(p_2)}-\expval{\mathcal{D^{C^{\max}}}(p_1)}}{p_2-p_1}}, p_2 > p_1.
\end{equation}
In case of phase damping channel,  \(\mathcal{R}^{C^{\max}}\) for the \(GHZ\)- and the \(W\)-class states are  $0.225$ and $0.0675$ respectively which is much lower than for the depolarising channel, $1.0$ and $1.052$. Hence higher decay rate dictates the channel to be the depolarising channel otherwise phase damping channel. On the other hand, opposite picture emerges when local ADC acts on the $GHZ$- and the $W$-class states. In particular, \(\mathcal{R}^{C^{\max}}<0\) for ADC. Therefore, the sign of \(\mathcal{R}^{C^{\max}}\) can be used to discriminate Pauli channels from ADC.

\small 
\begin{table}
\begin{tabular}{c}
Phase-flip Channel\\
\begin{tabular}{|c|c|c|c|c|}
\hline
$p=0.2$ & $p=0.4$ & $p=0.6$   \\
\hline
          \begin{tabular}{c|c|c }
            $N$ & $GHZ$ & $W$  \\ 
          \hline 
          $\langle \mathcal{D^{CD}} \rangle$ (p) & $0.514$  &  $0.298$ \\
          \hline
         $\sigma_\mathcal{D^{CD}}(p)$ & $0.182$  & $0.196$   \\
         \hline
         $\eta_\mathcal{D^{CD}}(p)$ &  $-3.630$ & $-2.738$   \\
         \end{tabular}         
         &    
           \begin{tabular}{c|c }
             $GHZ$ & $W$  \\ 
          \hline 
          $ 0.322 $   & $0.271$  \\
          \hline
          $0.154$   & $0.190$   \\
         \hline
          $1.554 $  & $0.861$  \\
         \end{tabular}              
         &        
           \begin{tabular}{c|c }
            $GHZ$ & $W$  \\ 
          \hline 
           $0.222$ & $0.254$  \\
          \hline
          $0.144$  & $0.186$   \\
         \hline
          $0.766$  &  $1.029$ \\
         \end{tabular}      \\
\hline          
\end{tabular}\\
Depolarizing Channel  \\
\begin{tabular}{|c|c|c|c|c|}
\hline
$p=0.2$ & $p=0.4$ & $p=0.6$  \\
\hline
          \begin{tabular}{c|c|c }
            $N$ & $GHZ$ & $W$  \\ 
          \hline 
          $\langle \mathcal{D^{CD}} \rangle$ (p) & $0.218$  & $0.093$ \\
          \hline
         $\sigma_\mathcal{D^{CD}}(p)$ & $0.074$  & $0.059$  \\
         \hline
         $\eta_\mathcal{D^{CD}}(p)$ & $-1.801$  & $0.613$  \\
         \end{tabular}         
         &     
           \begin{tabular}{c|c }
             $GHZ$ & $W$  \\ 
          \hline 
           $0.032$   & $0.014$ \\
          \hline
          $0.012$  &  $0.009$ \\
         \hline
          $-0.040$  &  $0.722$  \\
         \end{tabular}              
         &       
           \begin{tabular}{c|c }
             $GHZ$ & $W$  \\ 
          \hline 
           $0.001$ & $0.0004$ \\
          \hline
          $0.0004$   & $0.0003$  \\
         \hline
          $-0.026$  & $0.771$   \\
         \end{tabular}      \\
\hline                      
\end{tabular}\\
Amplitude-damping Channel\\
\begin{tabular}{|c|c|c|c|c|}
\hline
$p=0.2$ & $p=0.4$ & $p=0.6$  \\
\hline
          \begin{tabular}{c|c|c }
            $N$ & $GHZ$ & $W$  \\ 
          \hline 
          $\langle \mathcal{D^{CD}} \rangle$ (p) & $0.504$  & $0.256$ \\
          \hline
         $\sigma_\mathcal{D^{CD}}(p)$ &  $0.160$ & $0.156$  \\
         \hline
         $\eta_\mathcal{D^{CD}}(p)$ & $-6.365$   & $0.135$  \\
         \end{tabular}         
         &         
           \begin{tabular}{c|c }
             $GHZ$ & $W$  \\ 
          \hline 
           $0.261$  & $0.146$  \\
          \hline
          $0.088$  & $0.089$  \\
         \hline
          $0.758$  & $0.085$  \\
         \end{tabular}              
         &         
           \begin{tabular}{c|c }
             $GHZ$ & $W$  \\ 
          \hline 
           $0.115$  & $0.069$ \\
          \hline
          $0.040$  & $0.042$ \\
         \hline
          $0.577$  &  $0.078$ \\
         \end{tabular}      \\
\hline                                           
\end{tabular}\\

\end{tabular}
\caption{Statistical quantities  of the normalized frequency distributions of $\mathcal{D^{CD}}$ for \(GHZ\)- and \(W\)- class states. Other specifications are same as in Table \ref{tab:mean_noise}. }
\label{tab:GHZ-W-NFD-D}
\end{table}
\normalsize
   \begin{figure*}
		\includegraphics[width=\linewidth]{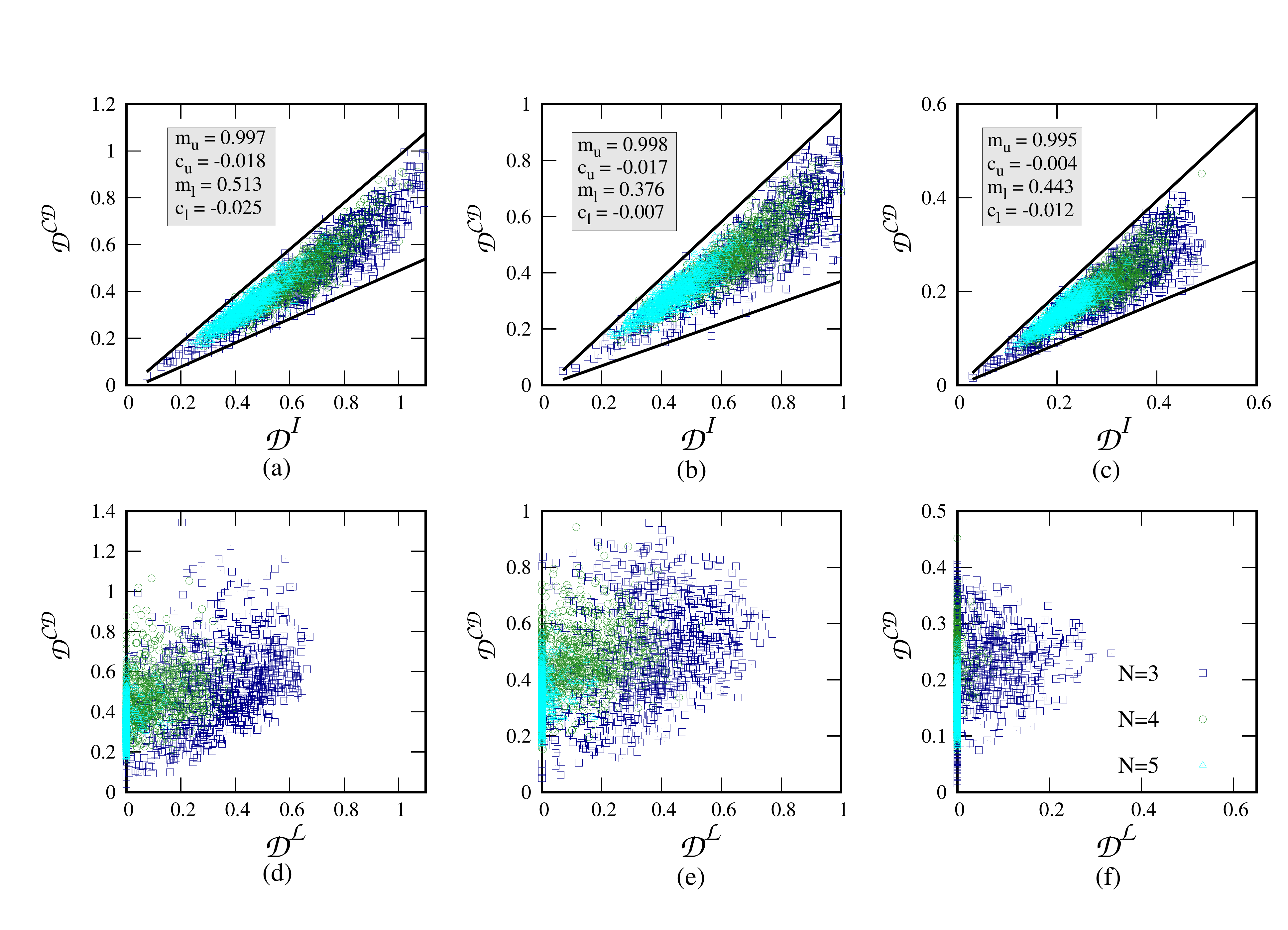}
		\caption{(a)-(c) Distributed classical discord $\mathcal{D^{CD}}$, (ordinate) against distributed mutual information, $\mathcal{D^{I}}$ (abscissa). (d)-(f). Behavior of $\mathcal{D^{CD}}$ (vertical axis) with respect to $\mathcal{D^{L}}$ (horizontal axis). Panel (a) and (d) represent local PDC, (b) and (e) are for local ADC while (c) and (f) indicate depolarizing channels.  In (a) - (c), slopes and constants of the boundary lines are mentioned in labels (see also Table \ref{tab:ub_lb}). All the axes are dimensionless.}
		\label{fig:mutual_logneg}
\end{figure*}

\section{Connecting classical correlations with other correlations under decoherence}
\label{sec:CCvsQCTC}
Up to now, we have investigated how CC measures of a given multipartite state behave with noise. As mentioned before, the motivation  is to identify the characteristics of CC against noisy environment which are either similar or different from QC and total correlations.  We have already reported that overall behavior of CC measures with noise especially  the axiomatic ones, does not show any difference with QC measures for randomly generated states.

Quantitatively, we will now establish a connection of CC measures with entanglement measures \cite{Horodecki_RMP_2009} like logarithmic negativity \cite{Werner_PRA_2002, Plenio_PRL_2005} and entanglement of formation \cite{wootters98} as well as total correlations, quantified via mutual information \cite{henderson2001, Winter_PRA_2005, Westmoreland_PRA_2006}, \(\mathcal{I}(\rho_{B_1B_2})=S(\rho_{B_1})+S(\rho_{B_2})-S(\rho_{B_1B_2})\). Specifically, we explore the connection between $\mathcal{D^{\chi}}$ where \(\chi \in\{C^{\max},\mathcal{CD},\mathcal{LW}\}\) and $\mathcal{D^{\chi^{QC}}}$ with \(\chi^{QC}\in\{ \mathcal{L},\mathcal{EOF}\}\) when Haar uniformly generated states are passed through channels. First of all, we want to examine $\mathcal{D^{\chi}}$ and $\mathcal{D^{\chi^{QC}}}$ or $\mathcal{D^{I}}$ by varying \(p\) and secondly, for a fixed \(p\), how the relation changes with the increase of number of parties.
Before presenting the results, let us state the following proposition.

\noindent\textbf{Proposition III.} \emph{In an arbitrary $N$-qubit state, $\rho_{B_1\ldots B_N}$, distributed classical correlation quantified via classical discord and distributed entanglement of formation is bounded above by entropy of the nodal party multiplied by $(N-1)$, i.e., \((N-1)S(\rho_{B_1})\).}

\begin{proof}
 For any arbitrary three-qubit states, \(\rho_{B_1B_iB_{i+1}}\), we know \cite{koashi2004}
\begin{equation}
    E_f(\rho_{B_1B_i})+C^{\leftarrow}(\rho_{B_1B_{i+1}})\le S(\rho_{B_1}),
\end{equation}
where \(E_f(\rho_{B_1B_i})\) denotes the entanglement of formation \cite{wootters98}, \(C^{\leftarrow}(\rho_{B_1B_{i+1}})\) represents the classical discord where measurement is performed on $B_{i+1}$ party. Here \(i=2,\ldots,N\) with $B_{N+1}=B_2$. Hence, we get (\(N-1\)) such inequalities, adding all of them, we obtain \(\mathcal{D^Q}+\mathcal{D^{CD}}\le (N-1)S(\rho_{B_1})\), thereby obtaining an upper bound between distributed quantum and classical correlations. Notice that, \(\mathcal{D^Q}\ge 0\) and hence \(\mathcal{D^{CD}}\le (N-1)S(\rho_{B_1})\). 
\end{proof}

\begin{figure*}
		\includegraphics[width=\linewidth]{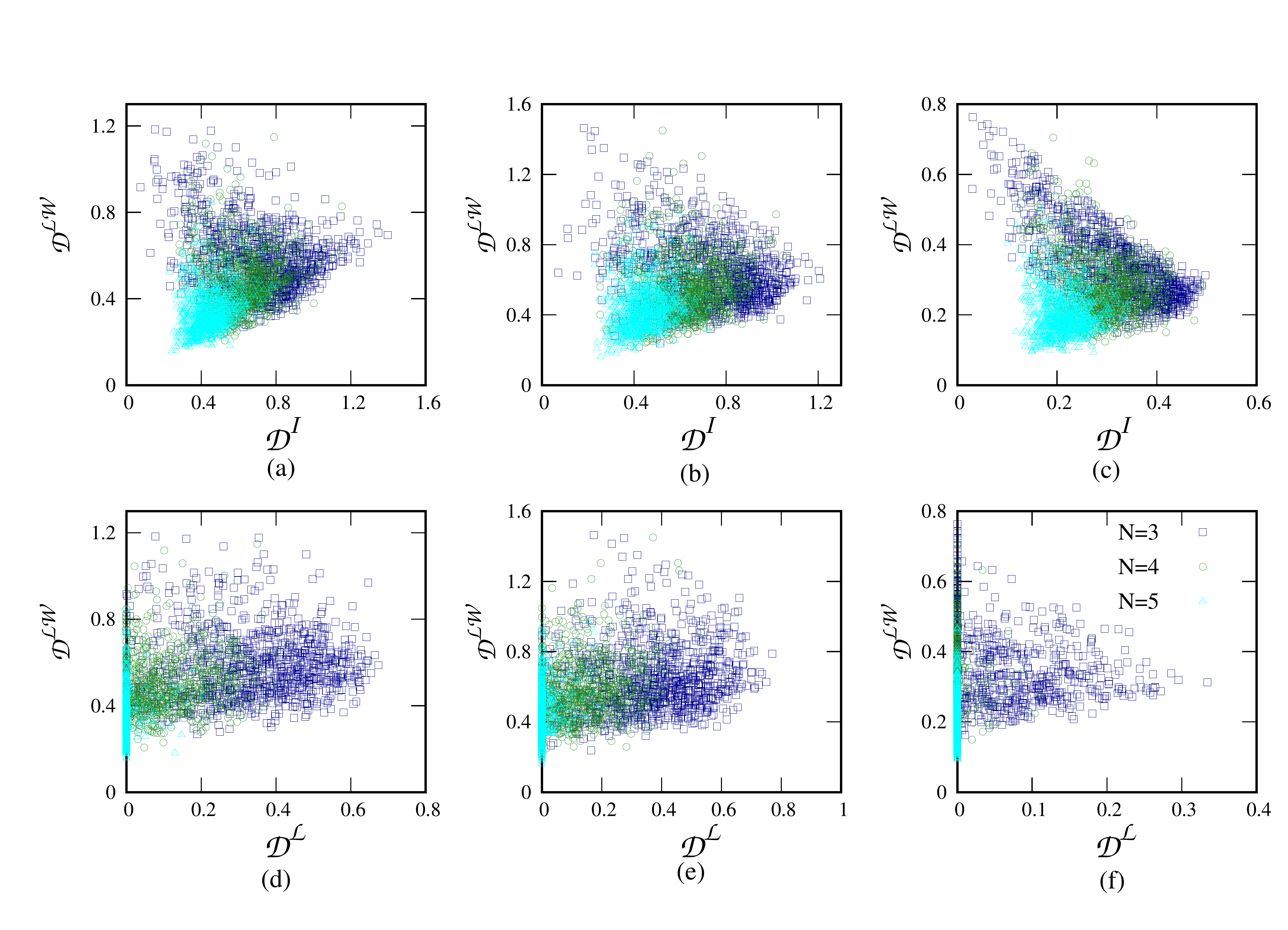}
		\caption{(a)-(c) Distributed local work, $\mathcal{D^{LW}}$ (ordinate) against distributed mutual information, $\mathcal{D^{I}}$ (abscissa). (d)-(f). Behavior of $\mathcal{D^{LW}}$ (vertical axis) with respect to $\mathcal{D^{L}}$ (horizontal axis).   All other specifications are same as in Fig. \ref{fig:mutual_logneg}.  All the axes are dimensionless.}
		\label{fig:mutual_local_work}
\end{figure*}

When different kinds of noise acts on individual qubits, we observe that for a given noise strength  \(p\) in the channel, for a fixed $\mathcal{D^I}$, \(\mathcal{D^{CD}}\in[l,u]\), where $l$ and $u$ represent lower and upper bounds of $\mathcal{D^{CD}}$. Interestingly, we find that with the increase of $\mathcal{D^I}$, the difference between $u$ and $l$, i.e., \(\Delta^{\mathcal{D^{CD}}}=(u-l)\) decreases with $\mathcal{D^I}$ which is true for all \(N\) and for any strength of noise, \(p\). We will show that these bounds can easily be obtained for any channels. 
Moreover, the nonvanishing region in the \((\mathcal{D^I}, \mathcal{D^{\chi}})\)- and  \((\mathcal{D^L}, \mathcal{D^{\chi}})\)-plane shift towards left with \(\chi\) being any measures of CC  and shrinks more and more as number of parties increases for a fixed \(p\) (see Figs. \ref{fig:mutual_logneg}, and  \ref{fig:mutual_local_work}). 

{\it CC vs. total correlations.} The relation between CD and mutual information is more direct. From the definition of quantum discord, we know that mutual information is linearly connected with classical correlation,  i.e., \(\mathcal{CD}^{\rightarrow}(\rho_{B_1B_i})=\mathcal{I}(\rho_{B_1B_i})- D(\rho_{B_1B_i}), \quad i\in\{2,\ldots,N\}\) where \(D\) is quantum discord. Of course, the definition also holds for \(\mathcal{CD}^{\rightarrow}\).  Adding the relation \(N-1\) times  for $N$-party states, we get \(\mathcal{D^{CD}}=\mathcal{D^I}-\mathcal{D}^D\). 

For Haar uniformly generated states, such linear relation becomes prominent. Moreover, Fig. \ref{fig:mutual_logneg} suggests that there exist an upper and a lower bounds of \(\mathcal{D^{CD}}\) which is linear with \(\mathcal{D^{I}}\). Hence, we can express the bound, which we describe as optimum value of classical discord, denoted by $\mathcal{D_\text{opt}^{CD}}$ with an equation for straight line, given by
 \begin{equation}
     \mathcal{D_\text{opt}^{CD}}=m\mathcal{D^{I}}+c,
     \label{eq::new_discord}
 \end{equation}
where \(m\) denotes the slope of the line and \(c\) is a constant. The  upper and lower lines  indicate the nonvanishing region in the \((\mathcal{D^I}, \mathcal{D^{CD}})\)-plane, and the corresponding slopes of these lines are referred to as \(m_{u}\) and \(m_{l}\) respectively while the constants are denoted as \(c_u\) and \(c_l\) respectively (see Table \ref{tab:ub_lb} for different noise, \(p\) and \(N\)). 

To determine the slope, we divide \(\mathcal{D^{I}}\) (the $x$-axis in Fig. \ref{fig:mutual_logneg}) into small segments, compute the maximum and minimum values of \(\mathcal{D^{CD}}\) in each such intervals and fit these values with a straight line which gives the slope, \(m\) and the constant $c$. Again we observe that with noise, the area of the triangle formed by two lines, \(\mathcal{D^{CD}}_{\max}\) and \(\mathcal{D^{CD}}_{\min}\) decreases with \(N\). The feature is independent of channels applied on the state. Such a linear relation with total correlation does not remain true for other CC measures. The region in the (\(\mathcal{D^{I}}, \mathcal{D^{CD}}\))-plane decreases with the increase of \(N\) like QC measures as shown in Fig. \ref{fig:mutual_logneg}. 
\small 
\begin{table} 
\begin{tabular}{c}
Phase-flip channel\\
\scalebox{0.90}{
\begin{tabular}{|c|c|c|}
\hline
$p=0.2$ & $p=0.4$ & $p=0.6$  \\
\hline
          \begin{tabular}{c|c|c|c}
            $N$  & $3$ & $4$ & $5$ \\ 
          \hline 
          $m_{u}$   &$0.996$  &  $0.965$ & $0.955$  \\
          \hline
         $m_{l}$   &$0.512$  &$0.509$ & $0.484$    \\
         
         \end{tabular}         
         &         
          \begin{tabular}{c|c|c}
           $3$ & $4$ & $5$ \\
          \hline 
           $0.983$  & $1.007$  & $1.01344$ \\
          \hline
          $0.527$ & $0.598$ & $0.784$   \\
         \end{tabular}             
         &         
          \begin{tabular}{c|c|c}
           $3$ & $4$ & $5$ \\
          \hline 
           $0.997$ & $0.98$ & $1.006$  \\
          \hline
           $0.584$ & $0.629$ & $0.968$ \\
         \end{tabular}  \\       
\hline          
\end{tabular}}\\
Depolarizing Channel \\
\scalebox{0.90}{
\begin{tabular}{|c|c|c|}
\hline 
$p=0.2$ & $p=0.4$ & $p=0.6$  \\
\hline 
          \begin{tabular}{c|c|c|c}
          $N$ & 3 & 4 & 5 \\ 
          \hline 
          $m_{u}$ & $0.994$   &$1.001$ &$0.945$  \\
          \hline
         $m_{l}$ & $0.443$  &$0.473$  & $0.476$  \\
         \end{tabular}         
         &         
          \begin{tabular}{c|c|c}
           3 & 4 & 5 \\ 
          \hline 
            $0.988$  & $0.986$  & $0.919$  \\
          \hline
           $0.407$ & $0.446$ & $0.634$  \\
         \end{tabular}             
         &         
          \begin{tabular}{c|c|c}
           3 & 4 & 5 \\ 
          \hline 
           $0.991$  & $0.961$ & $0.948$  \\
          \hline
          $0.363$& $0.399$ & $0.492$  \\
         \end{tabular}      \\   
\hline                      
\end{tabular}}\\
Amplitude-damping channel\\
\scalebox{0.90}{
\begin{tabular}{|c|c|c|} 
\hline 
$p=0.2$ & $p=0.4$ & $p=0.6$\\
\hline 
          \begin{tabular}{c|c|c|c}
          $N$ & 3 & 4 & 5 \\ 
          \hline 
          $m_{u}$ & $0.998$  & $0.990$ & $0.930$  \\
          \hline
         $m_{l}$ & $0.376$  &$0.497$  & $0.435$  \\
         \end{tabular}         
         &         
          \begin{tabular}{c|c|c}
           3 & 4 & 5 \\ 
          \hline 
           $0.983$ & $0.993$  & $0.882$  \\
          \hline
         $0.329$ & $0.441$ & $0.579$  \\
         \end{tabular}             
         &         
          \begin{tabular}{c|c|c}
           3 & 4 & 5 \\ 
          \hline 
           $0.886$ & $0.838$ & $0.767$  \\
          \hline
           $0.274$ & $0.367$ & $0.442$ \\
         \end{tabular}         \\
\hline                                           
\end{tabular}}\\

\end{tabular}
\caption{The slope of the straight lines enclosing the nonvanishing region in the \((\mathcal{D^I}, \mathcal{D^{CD}})\)-plane as shown in  Fig. \ref{fig:mutual_logneg} for different \(p\) and \(N\). 
}
\label{tab:ub_lb}
\end{table}
\normalsize


\section{conclusion}
\label{sec:conclusion}

The study of classical correlation (CC) from many perspectives is quite underexplored in the literature, even though it is important in many branches of physics, including quantum information theory, foundations of quantum mechanics, and condensed matter physics. However, due to the prominence as ingredients in numerous quantum protocols, quantum correlations (QCs) continue to be subjects of significant interest. QCs normally degrade in the presence of noise, however, after passing through noisy channels, CCs for classes of states either remain constant or get higher.
 
We studied the generic relation between the set of classical correlators (CCors) of the multipartite initial states and the noisy states for three different channels. We established a connection between the post-noise genuine and distributed CCors in terms of pre-noise CCors for the specific channel and found out that depending on the noise model, the dependence of CCors on the strength of noise changes from linear and quadratic to cubic ones, and in some specific direction, they remain  unaltered too. By  taking  two SLOCC inequivalent classes of states, the  \(GHZ\) and the \(W\) class states, we compute explicitly the relation between CCors of the initial and the final states and using such relation for the  \(W\)-class state, we proposed a protocol for identifying quantum channels. 

On numerous occasions, it has been demonstrated that generic Haar uniformly generated states can have some universal features that are very distinct from the qualities seen in a particular class of states. We confirm that this is also true for CC measures for generic states in presence of a noisy environment. In particular, we found the mean, standard deviation, and skewness of the distribution obtained from the randomly generated states for a fixed CC measure and  a fixed noise parameter in a given channel. We observed that the average CC diminishes as noise and the number of parties increases which is consistent with QC measures. The results remain unaffected by the CC measures chosen. In the case of three-qubits, we show that CC of generic states from the $W$-class  is more resilient to noise than that of the $GHZ$-class states. We exhibited that the noise affects equally the distribution of CC, QC, and total correlation measure which is quantified via mutual information in multipartite random states with the variation of noise. Furthermore, we found that, regardless of channels, mutual information and classical correlation in quantum discord are associated linearly. Our work shows that CCs, a crucial component of QC, can uncover some fascinating characteristics that can aid in the development of quantum processes.


\section*{acknowledgements}

We acknowledge the support from Interdisciplinary Cyber Physical Systems (ICPS) program of the Department of Science and Technology (DST), India, Grant No.: DST/ICPS/QuST/Theme- 1/2019/23. We  acknowledge the use of \href{https://github.com/titaschanda/QIClib}{QIClib} -- a modern C++ library for general purpose quantum information processing and quantum computing (\url{https://titaschanda.github.io/QIClib}) and cluster computing facility at Harish-Chandra Research Institute. 

\appendix


\section{Measures of Classical Correlations}
\label{cc_measures}
{\it Classical Correlators.}
Let us now characterize the two-site classical correlator present in any two-qubit state. Consider the Hilbert space $\mathcal{H} = C^2 \otimes C^2$. An arbitrary density matrix can be expressed as follows \cite{horodechki1995},
\begin{equation}
    \rho_{12}=\frac{1}{4} \left (\mathds{I} \otimes \mathds{I} + \vec{r} \cdot \vec{\sigma}  \otimes \mathds{I} + \mathds{I} \otimes \vec{s} \cdot \vec{\sigma} + \sum_{k,l = x,y,z} C'_{kl} \sigma_{k} \otimes \sigma_{l}\right),
\end{equation}
where $\mathds{I}$ stands for identity operator in two dimensional Hilbert space, $\vec{\sigma}\in\{\sigma_x,\sigma_y,\sigma_z\}$ are the standard Pauli matrices, magnetizations are referred by, $\{\vec{r},\vec{s}\}$ which are defined as \(r_k=\tr(\sigma_k\otimes\mathds{I}\rho_{12})\) and similarly \(s_k=\tr(\mathds{I}\otimes\sigma_k\rho_{12})\). The coefficients $C'_{kl}$, known as conventional classical correlator form a real matrix and they are written as
\begin{equation}
    C'_{kl} = \tr(\sigma_{k} \otimes \sigma_{l} \rho_{12}),\hspace{0.05in} k,l =x,y,z.
\end{equation}
 We find that \(-1\le C'_{kl} \le 1 \). It was shown that classical correlators play an important role to find out whether a state violates Bell inequality or not \cite{horodechki1995} or when a state is needful for quantum teleportation \cite{Horodecki_RMP_2009}. In similar fashion, one can write any arbitrary multipartite quantum state in forms of single-site , two-site \ldots , \(N\)-site CCors define in Eq. (\ref{correlator}).

{\it Axiomatic CC measures.} For a bipartite state, $\rho_{12}$, classical correlation measures were introduced which satisfy some desirable properties \cite{henderson2001}. We call them axiomatic CC measures. Prominent ones are classical correlation measure involved in the definition of quantum discord \cite{ollivier2001} which we call as classical discord \cite{henderson2001} and local work which is used to define  quantum correlation from the thermodynamic perspective \cite{horodecki2002}.

The classical correlation part of quantum discord of $\rho_{12}$ can be written as
\begin{equation}
   \mathcal{CD}^{\rightarrow}(\rho_{12}) = \underset{p_i}{\max}\left (S(\rho_1) -\sum_{i} p_i S(\rho_{i|1})\right )
\end{equation}
where $S(\rho)  = -\tr(\rho \log_2 \rho)$ represents the von Neumann entropy and $\rho_{i|1}$ is $\rho_{i|1}=\frac{\tr_2 (P_i\otimes \mathds{I} \rho_{12} P_i\otimes \mathds{I} )}{\tr (P_i\otimes \mathds{I}\rho_{12} \mathds{I} P_i\otimes \mathds{I} )}$, 
where $\{P_i\}$ is the rank-1 projective measurements over the first party, $p_i = \tr(P_i\otimes \mathds{I}\rho_{12}P_i\otimes \mathds{I} )$ and the maximization is performed over the set of  rank-1 projective measurements, $\{P_i\}$.  When measurement is performed on the second party, we denote the quantity as \(CD^{\leftarrow}(\rho_{12})\). 


\section{Different noisy channels}
\label{channels}

To analyse the effects of noise, we consider the situation where individual qubit of a multiparty system get affected due to the presence of environment. The local noise is characterized by dissipative as well as non-dissipative noise models. We consider well known noise models as phase damping channel, depolarising channel and amplitude damping channels. These noise model represented by Kraus operator formulation can be expressed as, $\rho_{in}\rightarrow\rho_{f}=\Lambda\left(\rho_{in}\right)$, where $\rho_{in}$ is the initial state and $\Lambda(.)$ is denoted by operator-sum representation, expressed as
 \(   \rho_f=\Lambda\left(\rho_{in}\right)=\sum_{r} K_{\mu} \rho_{in} K_{\mu}^\dagger\),
where \(\{K_{\mu}\}\) is the single qubit Kraus operator, following completeness relation, $\sum_{\mu} K_{\mu}^\dagger K_{\mu}=I$. The Kraus operators for PDC are given by
 \(K_{0} = \sqrt{1-\frac{p}{2}}\hspace{0.05in}\mathds{I}\), 
\(K_{1} = \sqrt{\frac{p}{2}} \sigma_z\).
and in case of depolarising channel,
    \(K_{0} = \sqrt{1-p}\hspace{0.05in}\mathds{I}\),
    \(K_{1} = \sqrt{\frac{p}{3}} \sigma_x\), \(K_{2} = \sqrt{\frac{p}{3}} \sigma_y\), and \(  K_{3} = \sqrt{\frac{p}{3}} \sigma_z\), 
while they are
\(K_{0} = \begin{pmatrix}
1 & 0 \\
0 & \sqrt{1-p}
\end{pmatrix}\)
\(K_{1} = \begin{pmatrix}
0 & \sqrt{p} \\
0 & 0
\end{pmatrix}\).
Here $p \, (0\le p\le 1)$ is the probability of applying a Kraus operator on a single qubit, and  it also represents the strength of noise.  Effects of noise on a $N$-party state computed as
\begin{equation}
 \rho_{in}^N\rightarrow\rho_f^N=\Lambda\left(\rho_{in}^N\right)=\sum_{\mu,\nu}\bigotimes_{i=1}^{N}K_{\mu}^{i} \rho_i^N {K_{\nu}^{i}}^\dagger,
\end{equation}
where $K_{\mu}$'s are set of Kraus operator for $N$ party state.

\section{CC measure for phase damping channel}

\label{cc_dephase}

We now describe how a $k$-th correlator for \(N\)-party state get affected due to the  phase damping channel. According to our convention, we put noise at each party, Kraus operator representation for phase damping channel is mentioned before. Generalizing  the scenario for a  \(N\)-party state, there are $2^N$ possible combination of Kraus operators, although it is easy to visualize that for fixed number of Kraus operators probability of acting noise remains same. This is expressed as below
\begin{eqnarray}
   K_r &=&\mathcal{P}\left( \underbrace{\ldots K_1\otimes K_1\ldots}_{r}\otimes\underbrace{\ldots K_0\otimes K_0\ldots}_{N-r}\right) \nonumber \\
   &\longrightarrow &  \left(\frac{p}{2}\right)^r\left(1-\frac{p}{2}\right)^{N-r},
\end{eqnarray}
where $\mathcal{P}$ defines different permutations of $r$ number of $K_1$ and $N-r$ number of $K_0$ Kraus operators. We denote all the permuted operator as single one, $K_r$.

The list of  formulas used to derive the action is as follows:
  (1) $K_i^{\dagger}=K_i$, (2)  $\{\sigma_i,\sigma_j\}=2\delta_{ij}$, where $\sigma_i$s are Pauli matrices, (3) $\tr(AB)=\tr(BA)$ and  (4) $\tr(\rho_{AB}\sigma_i\otimes \mathds{I})=\tr(\rho_A\sigma_i)$.
After transmission of each qubit of $\rho_{B_1\ldots B_N}$ through the phase damping channel, the final state is
\begin{equation}
    \rho_{B_1\ldots B_N}^f=\sum_{r=0}^{N}K_r\rho_{B_1\ldots B_N}. K_r^{\dagger}
\end{equation}
\begin{widetext}
Classical correlator of order $k$ of the final state reads as
\begin{eqnarray}
\nonumber C_{j_{1}\ldots j_{k}}' &=&\tr ( \sigma_{j_1} \otimes \ldots \otimes \sigma_{j_{k}} \rho_{B_1\ldots B_k}^f)\nonumber\\
&=&\tr ( \sigma_{j_1} \otimes \ldots \otimes \sigma_{j_{k}} \otimes \mathds{I} \rho_{B_1\ldots B_N}^f)\nonumber\\
&=&\tr ( \sigma_{j_1} \otimes \ldots \otimes \sigma_{j_{k}} \otimes \mathds{I} \sum_{r=0}^{N}K_r\rho_{B_1\ldots B_N} K_r^{\dagger})\nonumber\\
&=&\sum_{r=0}^{N}\tr ( \sigma_{j_1} \otimes \ldots \otimes \sigma_{j_{k}} \otimes \mathds{I} K_r\rho_{B_1\ldots B_N} K_r^{\dagger})\nonumber\\
&=&\sum_{r=0}^{N}\left(\frac{p}{2}\right)^r\left(1-\frac{p}{2}\right)^{N-r}\tr \{ \sigma_{j_1} \otimes \ldots \otimes \sigma_{j_{k}} \otimes \mathds{I} \mathcal{P}\left( \underbrace{\ldots \sigma_z\otimes \sigma_z\ldots}_{r}\underbrace{\ldots I \otimes I \ldots}_{N-r}\right)\rho_{B_1\ldots B_N}\nonumber\\
&\times& \mathcal{P}\left( \underbrace{\ldots \sigma_z\otimes \sigma_z \ldots}_{r}\otimes\underbrace{\ldots I \otimes I\ldots}_{N-r}\right)\}\nonumber\\
&=&\sum_{r=0}^{N}\left [\sum_{q=0}^{\min\{r,k\}}(-1)^q\binom{k}{q}\binom{N-k}{r-q}\right ]\left(\frac{p}{2}\right )^r \left (1-\frac{p}{2}\right )^{N-r}\tr(\sigma_{j_1} \otimes \ldots \otimes \sigma_{j_{k}}\rho_{B_1\ldots B_N})\nonumber.
\label{finalcorrelator}
\end{eqnarray}
\end{widetext}
In the last line, we use properties of trace and Pauli matrices. For $k$-correlator,  negative sign arises due to the non-commuting properties of Pauli matrices. 

\bibliographystyle{apsrev4-1}
\bibliography{bib}
\end{document}